\documentclass[twocolumn,english,groupedaddress, superscriptaddress]{revtex4-1}
\usepackage[T1]{fontenc}
\usepackage[latin9]{inputenc}
\usepackage{amstext}
\usepackage{graphicx}
\usepackage{wasysym}
\usepackage{babel}
\usepackage{color}
\usepackage[normalem]{ulem}

\begin{document}

\title{Dynamically induced cascading failures in power grids}

\author{Benjamin Schäfer}

\email{benjamin.schaefer@tu-dresden.de}
\affiliation{Chair for Network Dynamics, Center for Advancing Electronics Dresden
(cfaed) and Institute for Theoretical Physics, Technical University
of Dresden, 01062 Dresden, Germany}

\affiliation{Network Dynamics, Max Planck Institute for Dynamics and Self-Organization
(MPIDS), 37077 Göttingen, Germany}

\author{Dirk Witthaut }

\affiliation{Forschungszentrum Jülich, Institute for Energy and Climate Research
- Systems Analysis and Technology Evaluation (IEK-STE), 52428 Jülich,
Germany}

\affiliation{Institute for Theoretical Physics, University of Cologne, 50937 Köln,
Germany}

\author{Marc Timme}

\email{marc.timme@tu-dresden.de}
\affiliation{Chair for Network Dynamics, Center for Advancing Electronics Dresden
(cfaed) and Institute for Theoretical Physics, Technical University
of Dresden, 01062 Dresden, Germany}

\affiliation{Network Dynamics, Max Planck Institute for Dynamics and Self-Organization
(MPIDS), 37077 Göttingen, Germany}

\author{Vito Latora}

  \email{v.latora@qmul.ac.uk}

\affiliation{School of Mathematical Sciences, Queen Mary University of London,
London E1 4NS, United Kingdom}

\affiliation{Dipartimento di Fisica ed Astronomia, Universit\`a di Catania
  and INFN, I-95123 Catania, Italy}
  
\begin{abstract}
\section*{Abstract}
Reliable functioning of infrastructure networks is essential for our modern society. Cascading failures are the cause of most large-scale network outages. Although cascading failures often exhibit dynamical transients, the modeling of cascades has so far mainly focused on the analysis of sequences of steady states. In this article, we focus on electrical transmission networks and introduce a framework that takes into 
account both the event-based nature of cascades and the essentials of the network dynamics. We find that transients of the order of seconds in the flows of a power grid play a crucial role in the emergence of collective behaviors. We finally propose a forecasting method to identify critical lines and components in advance or during operation. Overall, our work highlights the relevance of dynamically induced failures on the synchronization dynamics of national power grids of different European countries and provides methods to predict and model cascading failures. 
\end{abstract}
\maketitle

\section*{Introduction}
Our daily lives heavily depend on the functioning of many natural and
man-made networks, ranging from neuronal and gene regulatory networks to communication systems, transportation
networks and electrical power
grids \cite{Newman2010,Latora2017}. Understanding the robustness of these
networks with respect to random failures and to targeted attacks is of
outmost importance for preventing system outages with severe
implications \cite{Albert2000}. Recent examples, as the
2003 blackout in the Northeastern United States \cite{USBlackout2003}, the major European
blackout in 2006 \cite{UCTE_Report2007} or the Indian blackout in 2012 \cite{CERC2012}, 
have shown that initially local and small events can trigger large area outages of
electric supply networks affecting millions of people, with
severe economic and political consequences \cite{Bialek2007}. Cascading events will become more likely in the future due to increasing load \cite{Pesc14} and additional fluctuations in the grid \cite{Schaefer2018}.
For this reason cascading failures have been studied intensively in statistical physics, 
and different network topologies and 
non-local effects have been considered and analyzed 
\cite{Simonsen2008,Hines2010,Brum13,Pahwa2014,Witthaut2015,
Plietzsch2016,Rohden2016,Manik2017,Ronellenfitsch2017}. 
Complementary studies have employed simplified topologies that admit
analytical insights, for instance in terms of percolation theory \cite{Callaway2000}
or minimum coupling \cite{Lozano2012}. Results have shown, for instance, the robustness 
of scale-free networks \cite{Albert2000,Albert2004,Boccaletti2006}, 
or the vulnerability of multiplex networks \cite{Buldyrev2010, Boccaletti2014,Battiston2014,Scala2016}.

Although real-world cascades often include dynamical
transients of grid frequency and flow with very well defined spatio-temporal structures, 
so far models of cascading failures have mainly focused on
event-triggered sequences of steady states
\cite{Witthaut2015,Plietzsch2016,Rohden2016,Crucitti2004,Crucitti2004a,Kinney2005,Ji2016,Pahwa2014}
or on purely dynamical descriptions of desynchronization without considering secondary failure of lines \cite{12powergrid,13powerlong,14bifurcation,Nish15,16critical}. In particular,
in supply networks such as electric power grids, which 
are considered as uniquely critical among all
infrastructures \cite{Kund94}, the failure of transmission lines 
during a blackout is determined not only by the network topology and 
by the static distribution of the electricity flow,  but also by the 
collective transient dynamics of the entire 
system. Indeed, during the severe outages mentioned 
above, cascading failures in electric power grids happened on 
time scales of dozens of minutes overall, but often started due to the failure of a single element \cite{Pourbeik2006}. Conversely, 
sequences of individual line overloads took place on a much shorter time scale 
of seconds \cite{UCTE_Report2007,USBlackout2003}, the time scale of
systemic instabilities, emphasizing the role of transient dynamics in the 
emergence of collective behaviors. For example, during the European blackout in 2006, a total of 33 high voltage transmission lines tripped within a time period of 1 minute and 20 seconds, with 30 of those lines failing within the first 19 seconds \cite{UCTE_Report2007}. Notwithstanding the importance of these fast transients, 
the causes, triggers and propagation of cascades induced by transient dynamics 
have been considered only in a few works \cite{Simonsen2008, Yang2017b}, and still need to be systematically 
studied  \cite{Bienstock2011}. Hence, 
we here focus on characterizing the dynamics of events that take place at the 
short time scale of seconds, which substantially contribute to the overall outages occurring in real grids.

This work complements the existing studies on cascading failures 
in power grids by linking nonlinear transient dynamics on short time scales to cascade events and simultaneously capturing line failures due to static overload. 
It is yet unrealistic to capture all aspects and time scales 
within a single model that is analytically tractable and provides mechanistic insights.
Most of the previous studies \cite{Witthaut2015,Plietzsch2016,Rohden2016,Crucitti2004,
Crucitti2004a,Kinney2005,Ji2016,Pahwa2014,Yang2017a}, based on
the analysis of sequences of steady states, consider the effects of
power plant shutdown or line outages and did not take into account any transient dynamical effects at all.
In contrast, a dynamical model might provide insights into cascading failures potentially induced on short time scales, thereby  characterizing the 
time scales relevant to the majority of line failures.

In this article, we propose a general framework to analyze the impact
of transient dynamics (of the order of seconds) on the outcome of cascading failures taking
place over a complex network. Specifically, we go beyond purely topological
or event-based investigations and present a dynamical model for electrical transmission networks
that incorporates both the event-based nature of cascades and the properties 
of network dynamics, including transients, which, as we will show, can
significantly increase the vulnerability of a network
\cite{Simonsen2008}. These transients describe the dynamical response of system variables, such as grid frequency and power flow, when one steady state is lost and the grid changes to a new steady state.
Combining microscopic nonlinear dynamics techniques 
with a macroscopic statistical analysis of the system, we will first
show that, even when a network seems to be robust because in the large
majority of the cases the initial failure of its lines does only have
local effects, there exist a few specific lines which can trigger
large-scale cascades. We will then analyze the vulnerability of a network by looking at the dynamical properties of cascading failures.   
To identify the critical lines of the network we introduce and analytically
derive a flow-based classifier that is shown to outperform measures
solely relying on the network topology, local loads
or network susceptibilities (line outage distribution factors). Finally,
we  find that the distance of a line failure from the initial trigger and the time of the line failure are highly correlated, especially
when   a measure of  effective distance is adopted \cite{Brockmann2013}.


\begin{figure*}
\begin{centering}
\includegraphics[width=2\columnwidth]{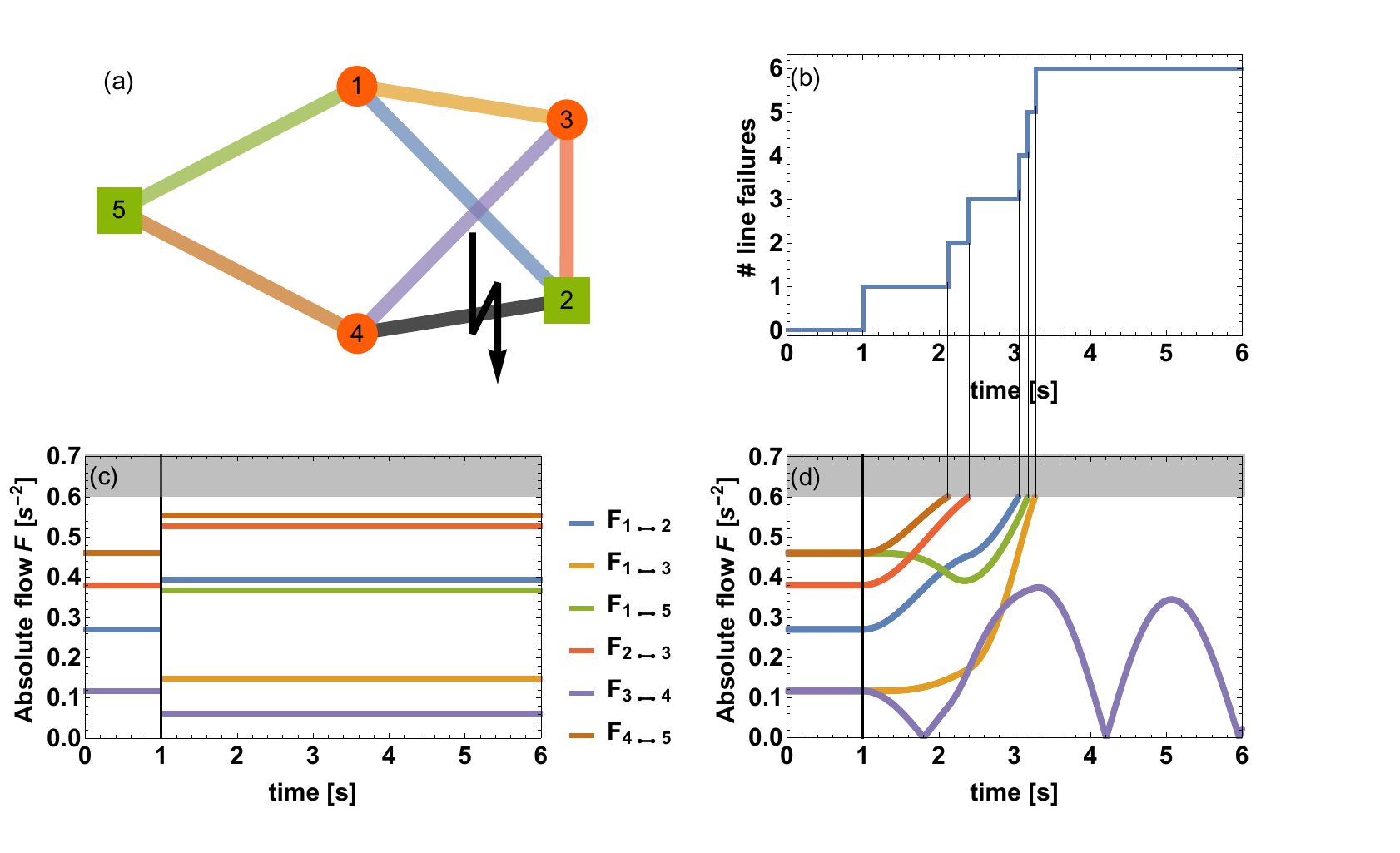}
\par\end{centering}

\caption{Dynamical overload reveals additional line failures compared to
static flow analysis. (a) A  five node power network with
two generators $P^+=1.5/s^2$ (green squares), three consumers $P^-=-1/s^2$ (red circles),
homogeneous coupling $K\approx1.63/s^2$, and tolerance $\alpha=0.6$ 
is analyzed. To trigger a cascade, we remove the line marked with a
lightening bolt (2,4) at time $t=1 s$. Other lines are color coded as the flows in (c) and (d).
(b) We observe a cascading failure with several additional
line failures after the initial trigger due to the propagation of overloads. 
(c) The common quasi-static approach of analyzing fixed point flows would
have predicted no additional line failures, since the new fixed point is
stable with all flows below the capacity threshold.
(d) Conversely, the transient dynamics from the initial to the
new fixed point overloads additional lines which then fail 
when their flows exceed their capacity (gray
area).
\label{fig:Example of dynamical overload}\protect \\
}
\end{figure*}

\section*{Results}
\subsection*{The dynamics of cascading failures}
\label{sec:ElementaryCascades}

Failures are common in many interconnected systems, such as
communication, transport and supply networks, which are fundamental
ingredients of our modern societies.  Usually, the failure of a
single unit, or of a part of a network, is modeled by removing or
deactivating a set of nodes or lines (or links) in the corresponding
graph \cite{05vuln}.  The most elementary damage to a network consists
in the removal of a single line, since removing a node is equivalent
to deactivate more than one line, namely all those lines incident in
the node.  For this reason, in the following of this work, we
concentrate on line failures.  In practice, the malfunctioning of a
line in a transportation/communication network can either be due to an
exogenous or to an endogenous event
\cite{PhysRevLett.93.068701,08Meloni}.  In the first case, the line
breakdown is caused by something external to the network. Examples are
the lightning strike of a transmission line of the electric power
grid, or the sagging of a line in the heat of the summer. In the case
of endogenous events, instead, a line can fail because of an overload
due to an anomalous distributions of the flows over the
network. Hence, the failure is an effect of the entire network.
\\ Complex networks are also prone to cascading failures. In
these events, the failure of a component triggers the successive
failures of other parts of the network. In this way, an initial local
shock produces a sequence of multiple failures in a domino mechanism
which may finally affect a substantial part of the network. Cascading
failures occur in transportation systems
\cite{ccolak2016understanding,lima2016understanding}, in computer
networks \cite{Echenique2005}, in financial systems
\cite{Petrone2016}, but also in supply networks \cite{Buldyrev2010}.
When, for some either exogenous or endogenous reason, a line of a
supply network fails, its load has to be somehow redistributed to the
neighboring lines. Although these lines are in general capable of handling
their extra traffic, in a few unfortunate cases they will also go
overload and will need to redistribute their increased load to their
neighbors. This mechanism can lead to a cascade of failures, with a
large number of transmission lines affected and malfunctioning at the
same time. One particular critical supply network is the electrical
power grid displaying for example large-scale cascading failures
during the blackout on 14 August 2003, affecting millions of people in
North America, and the European blackout that occurred on 4 November
2006.
%
In order to model cascading failures in power transmission networks,
we propose to use the framework of the swing equation, see Eq.~(\ref{eq:Swing equation}) in Methods,  to evaluate, at each time, the actual
power flow along the transmission lines of the network and compare it
to the actual available capacity of the lines, see Methods.
Typical studies of network robustness and cascading failures in power
grids adopted quasi-static perspectives \cite{Witthaut2015,Plietzsch2016,Rohden2016,Crucitti2004,Crucitti2004a,Kinney2005,Ji2016,Pahwa2014}
based on fixed-point estimates of the variables 
describing the node states. Such approach, in the context of the
swing equation, is equivalent to the evaluation of the angles
$\{ \theta_{i} \}$ as the fixed point solution of Eq.~(\ref{eq:Swing equation})
or power flow analysis \cite{Kund94}.
In contrast, we use here the swing
equation to dynamically update the angles $\theta_{i}\left(t\right)$
as functions of time, and to compute real-time estimates of the flow on
each line. The flow on the line $(i,j)$ 
at time $t$ is obtained as:  
\begin{equation}
F_{ij}\left(t\right)=K_{ij}\sin\left(\theta_{j}\left(t\right)-\theta_{i}\left(t\right)\right).
  \label{eq:dyn_flows}
\end{equation}
Having the time evolution of the flow along the line $(i,j)$, we  
compare it to the capacity $C_{ij}$ of the line, i.e., to the maximum flow
that the line can tolerate.  There are multiple options how 
we can define the capacity of a line in the framework of 
the swing equation. One possibility is the following. 
The dynamical model of Eq.~(\ref{eq:Swing equation}) itself 
would allow a maximum flow equal
to $F_{ij}=K_{ij}$ on the line $(i,j)$
. However, in realistic settings, ohmic
losses would induce overheating of the lines which has to be avoided.
Hence, we assume that the capacity $C_{ij}$ is set to be a
tunable percentage of $K_{ij}$. In order to prevent damage and
keep ground clearance \cite{Wood13,Mach08}, 
the line $(i,j)$
 is then shut down if the flow on 
it exceeds the value $\alpha K_{ij}$, where 
$\alpha\in\left[0,1\right]$ is
 a control parameter of the model.
The overload condition on the line $(i,j)$
 at time $t$ 
finally reads: 
\begin{equation}
\text{overload:    }    \left|F_{ij}(t)\right|>C_{ij}=\alpha K_{ij}.
  \label{eq:Capacity rule}
\end{equation}
Notice that the capacity $C_{ij}=\alpha K_{ij}$ is an absolute capacity, i.e.,
it is independent from the initial state of the system. This is
different from the definition of a relative
capacity,  
$\tilde{C}_{ij}:=\left(1+\alpha\right)F_{ij}\left(0\right)$, 
which has been commonly adopted in the literature 
\cite{Motter2002,Crucitti2004,Simonsen2008}.

  Having defined the fixed point of the grid, given by the
  solution of Eq.~(\ref{eq:stable state}),
  and the capacity of each line, we explore the 
  robustness of the network with respect to line failures.
  We first consider the ideal scenario in which all elements of
  the grid are working properly, i.e., all
  generators are running as scheduled and all lines are operational.
  We say that the grid is $N-0$ stable \cite{Koc2014}   
  if the network has a stable fixed point and the flows
  on all lines are within the bounds of the security limits, i.e. do not violate the overload condition
  Eq.~(\ref{eq:Capacity rule}), where the flows are
  calculated by inserting the fixed point solution into
  Eq.~(\ref{eq:dyn_flows}). 

  Next, we assume the initial failure of a single transmission line. 
  We call the new network in which the corresponding line has been
  removed the $N-1$ grid. 
  Since the affected transmission line can be any of the $|E|$ lines of the
  network, we have $|E|$ different $N-1$ grids. 
 If the $N-1$ grid still has a fixed
  point for all possible $|E|$ different initial failures,
  and all of these fixed points result in flows within the capacity
  limits, the grid is said to be 
  $N-1$ stable \cite{Mach08,Kund94,Wood13}.
  While traditional cascade approaches usually test
  $N-0$ or $N-1$ stability
  using mainly static flows, our proposal is to investigate cascades
  by means of dynamically updated flows according to the  
  power grid dynamics of Eq.~(\ref{eq:dyn_flows}). This allows 
  for a more realistic modeling of real-time overloads and
  line failures. In practice, this means to solve the swing equation 
  dynamically, update flows and compare to the capacity rule
  Eq.~(\ref{eq:Capacity rule}), removing lines whenever they exceed
  their capacity. Thereby, our $N-1$ stability criterion demands not
  only the stable states to stay within the capacity limits but also
  includes the transient flows on all lines. See Supplementary Note 1
  for details on our procedure, and Supplementary Note 6 for an investigation of the case of lines tripping 
  after a finite overload time.

\begin{figure*}
\begin{centering}
\includegraphics[width=2\columnwidth]{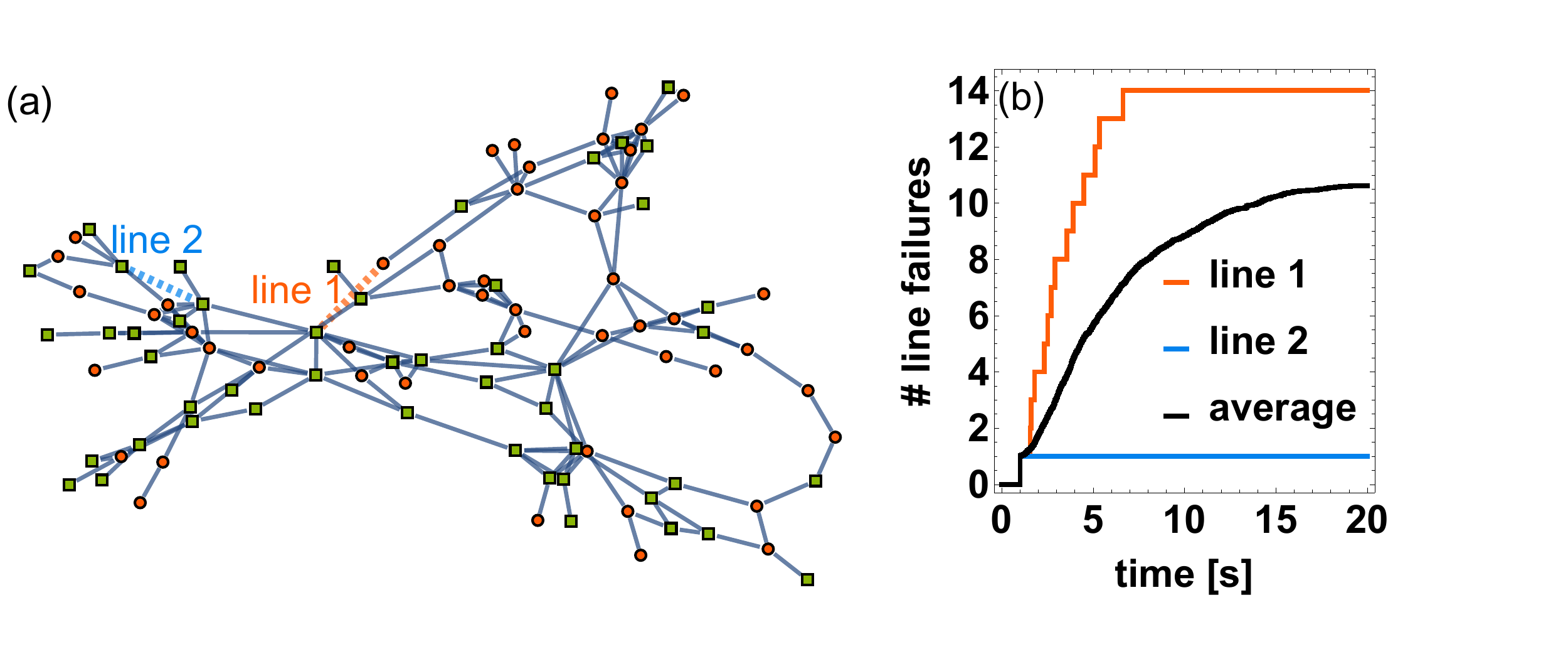}
\par\end{centering}
\caption{The effect of a cascade of failures strongly depends on the choice of the initially
  damaged line. (a) The network of the Spanish power grid with 
distributed generators with $P^+=1/s^{2}$ (green 
squares) and consumers with $P^-=-1/s^{2}$ (red circles), 
homogeneous coupling $K=5/s^{2}$, and tolerance $\alpha=0.52$ 
is analyzed. Two different trigger lines are selected. 
(b) The number of line failures as a function of time for the two different trigger lines highlighted in panel (a) and for an average over all possible
initial damages. 
Some lines do only cause a single line failure, while others affect
a substantial amount of the network. On average most line failures do
take place within the first $\approx 20$ seconds of the cascade.\label{fig:Spanish map and single line cuts} }
\end{figure*}

In order to illustrate how our dynamical model for cascading failures
works in practice, we first consider the case of the network with
$N=5$ nodes and $|E|=7$ lines shown in Fig.~\ref{fig:Example of
  dynamical overload}.  We assume that the network has two generators,
the two nodes reported as green squares, characterized by a positive
power $P^+=1.5/s^{2}$, and three consumers, reported as red circles,
with power $P^{-}= - 1/s^{2}$. For simplicity we have adopted here
  a modified ``per unit system'' obtained by replacing real machine parameters
  with dimensionless multiples with respect to reference values.
  For instance, here  a ``per unit'' mechanical power $P_{\text{per unit}}=1/s^2$
  corresponds to the real value 
$P_{\text{real}}=100MW$
\cite{Mach08,Kund94}. Moreover, we assume homogeneous line parameters
throughout the grid, namely, we fix the coupling for each couple of
nodes $i$ and $j$ as $K_{ij}= K a_{ij}$, with $K = 1.63/s^{2}$. In
order to prepare the system in its stable state, we solve
Eq.~(\ref{eq:stable state}) and calculate the corresponding flows at
equilibrium. We then fix a threshold value of $\alpha=0.6$. With such
a value of the threshold, none of the flows is in the overload
condition of Eq.~(\ref{eq:Capacity rule}), and the grid is $N-0$
stable.  Next, we perturb the stable steady state of the grid with an
initial exogenous perturbation.  Namely, we assume that line $(2,4)$
fails at time $t=1$, due to an external disturbance. By using again
the static approach of Eq.~(\ref{eq:stable state}) to calculate the
new steady state of the system, it is found that all flows have
changed but they still are all below the limit of 0.6, as shown in
Fig.~\ref{fig:Example of dynamical overload}(c). Hence, with respect
to a static analysis, the grid is $N-1$-stable to the failure of line
$(2,4)$.  Despite this, the capacity criterion in Eq.~(\ref{eq:Capacity
  rule}) can be violated transiently, and secondary outages emerge
dynamically. As Fig.~\ref{fig:Example of dynamical overload}(d) shows,
this is indeed what happens in the example considered.  Approximately
one second after the initial failure, the line $(4,5)$ is overloaded,
which causes a secondary failure, leading to additional overloads on
other lines and their failure in a cascading process that eventually leads to the disconnection of the entire grid. The whole
dynamics of the cascade of failures induced by the initial removal of
line $(2,4)$ is reported in Fig.~\ref{fig:Example of dynamical
  overload}(d). A dynamical update of the cascading algorithm is also shown in Supplementary Movie 1.

Dynamical cascades are not limited to small networks as the one 
considered in this example, but also appear in large networks.
In order to show this, we have implemented our model for cascading
failures on a network based on the real structure of the Spanish
high voltage transmission grid. The network is reported  
in Fig.~\ref{fig:Spanish map and single line cuts} and has
$N_\text{Spanish}=98$ nodes and
$\left| E \right|_\text{Spanish}=175$ edges.   
We remark that, while we have complete 
knowledge of the network topology, due to only partial information available on 
line parameters and power distribution, we have to estimate those missing parameters.
Therefore, we have investigated several different power distributions and coupling scenarios in our simulations, 
including homogeneous versus heterogeneous coupling, as well as considering 
cases with many small power plants, compared to cases of fewer but larger plants.
All parameter choices we have adopted are further specified in Supplementary Note 1 and the Data Availability Statement.
We start by selecting a set of distributed generators (green squares),
each with a positive power $P^+=1/s^{2}$, and consumers
(red circles), with negative power $P^-=-1/s^{2}$.
As in the case of the previous example, we adopt a 
homogeneous coupling, namely we fix $K_{ij}= K  a_{ij}$
with $K=5/s^{2}$ for each couple of nodes $i$ and $j$. We also
fix a tolerance value $\alpha=0.52$, such that 
none of the flows is in the overload condition of Eq.~(\ref{eq:Capacity rule}), and the
grid is initially $N-0$ stable. We notice from the effects of
cascading failures shown in Fig.~\ref{fig:Spanish map and single line cuts}
that the choice of the trigger line significantly 
influences the total number of lines damaged during a cascade. 
For instance, the initial damage of line 1 (dashed red line) causes a large cascade of
failures with 14 lines damaged in the first seven
seconds, while the initial damage of line 2 (dashed blue line)
does not cause any further line failure, as the initial shock is in this case perfectly
absorbed by the network. Figure \ref{fig:Spanish map and single line cuts} also displays the 
average number of failing lines as a function of time. Here, we average
over all lines of the network considered as initially damaged
lines. We notice that the cascading process is relatively fast, with all 
failures taking place within the first
$T_{\text{Cascade}}=20\,\text{s}$.
This further supports the adoption of the swing equation,
which is indeed mainly used to describe short time
scales, while more complex and less tractable models are required
to model longer times \cite{Mach08}.

\subsection*{Statistics of Dynamical Cascades}
\label{sec:StatisticsOfCascades}

\begin{figure*}[ht]
\begin{centering}
\includegraphics[width=2\columnwidth]{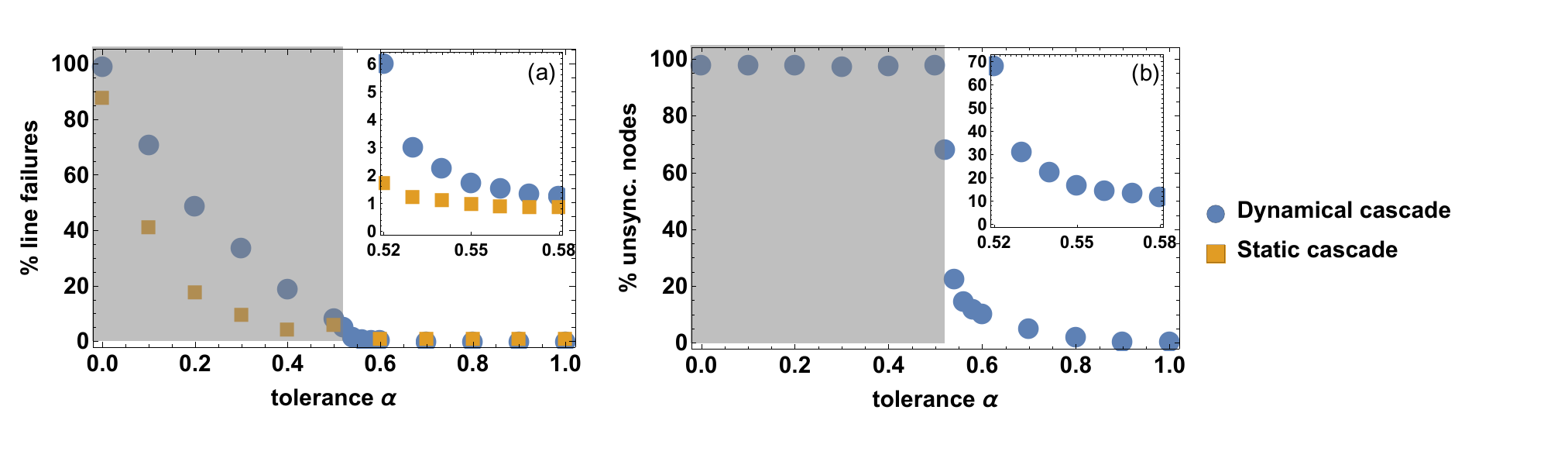}
\par\end{centering}
\caption{Effects of cascading failures in the Spanish power grid
  under different levels of tolerance.  (a) The percentage of line
  failures in our model of cascading failures (circles), under different
  values of tolerance $\alpha$ is compared to the results of a static
  fixed point flow analysis (squares). The static analysis largely
  underestimates the actual number of line failures in a dynamical
  approach. The difference between static and dynamical analysis is
  especially clear in the inset where we focus on the lowest values of
  $\alpha$ at which the network is $N-0$ stable. The gray area is
  $N-0$ unstable, i.e., the network without any external damage
  already has overloaded lines. (b) Percentage
  of unsynchronized (damaged) nodes after the cascade as a function of
  the tolerance $\alpha$.  All analysis has been performed under the
  same distribution of generators and consumers as in
  Fig.~\ref{fig:Spanish map and single line cuts}, with homogeneous
  coupling of $K=5 s^{-2}$.
\label{fig:Spanish nodes and links vs tolerance}\protect
}
\end{figure*}

To better characterize the potential effects of cascading failures in
electric power grids, we have studied the statistical properties of
cascades on the topology of real-world power transmission grids, such
as those of Spain and France \cite{Rosato2007}. 
In particular, we have considered the two systems under different values
of the tolerance parameter $\alpha$ \cite{Crucitti2004}, and for
various distributions of generators and consumers on the network. As in
the examples of the previous section, we have also analyzed all
possible initial damages triggering the cascade.  To assess the
consequences of a cascade, we have focused on the following two
quantities.  First, we analyze the number of lines that suffered an
overload, and are thus shut down during the cascading failure process.
This number is a measure of the total damage suffered by the system
in terms of loss of its connectivity.  
Second, we record the fraction of nodes that have experienced a
desynchronization during the cascade,
which represents a proxy for the number of consumers
affected by a blackout, see
Supplementary Note 1 for details on the implementation. In both the cases of affected lines and 
affected nodes, the numbers we look at are those obtained at the
end of the cascading failure process.

Fig.~\ref{fig:Spanish nodes and links vs tolerance} shows the results
obtained for the case of the network of the Spanish power transmission
grid. The same homogeneous coupling and distribution of generators and
consumers is adopted as in Fig.~\ref{fig:Spanish map and single line cuts}.  We have considered each of the lines as a possible initial trigger
of the cascade, and averaged the final number of line failures and
unsynchronized nodes over all realizations of the dynamical process. We have repeated this for multiple values of the tolerance
coefficient $\alpha$. As expected, a larger tolerance results in fewer
line failures and fewer unsynchronized nodes, because it makes the 
overload condition of Eq.~(\ref{eq:Capacity rule}) more difficult to be satisfied.
As we decrease the network tolerance $\alpha$, the total number of
affected lines and unsynchronized nodes after the cascade suddenly
increases at a value $\alpha\approx0.5$, where we start to observe a
propagation of the cascade induced by the initial external damage.
Crucially, a dynamical approach, as the one considered in our model,
identifies a significantly larger number of line failures (circles)
compared to a static approach (squares). This is clearly visible in
the inset of the left hand side of
Fig.~\ref{fig:Spanish nodes and links vs tolerance}, where we zoom
to the lowest values of $\alpha$ at which 
the network is $N-0$ stable. For instance, at $\alpha=0.52$ our model predicts that 
an average of six lines of the Spanish power grid are affected
by the initial damage of a line of the network through a propagation of failures.
Such a vulnerability of the network is completely unnoticeable by 
a static approach to cascading failures based on the analysis of fixed points.
The static approach reveals in fact that on average only another line of the
network will be affected. 
We also note that the increase in the number 
of unsynchronized nodes for decreasing values of $\alpha$ is much sharper
than that for overloaded lines.
 Below a value of
$\alpha\approx0.5$ the number of unsynchronized nodes jumps to 
$100\%$. This transition  indicates a loss of the $N-0$ stability of the system, meaning that, 
already in the unperturbed state several lines are overloaded according to the
capacity criterion in Eq.~(\ref{eq:Capacity rule}) and thus fail.
To study only genuine effects of cascades, in the following we 
restrict ourselves to the case $\alpha>0.5$, where the grid is $N-0$ stable,
but not necessarily $N-1$ stable. Furthermore, to assess the final
impact of a cascade on a network, we mainly focus on total number of affected
lines \cite{USBlackout2003,UCTE_Report2007}. As discussed in the last section, damages to lines are indeed the most elementary
type of network damages.

\begin{figure*}
\begin{centering}
\includegraphics[width=2\columnwidth]{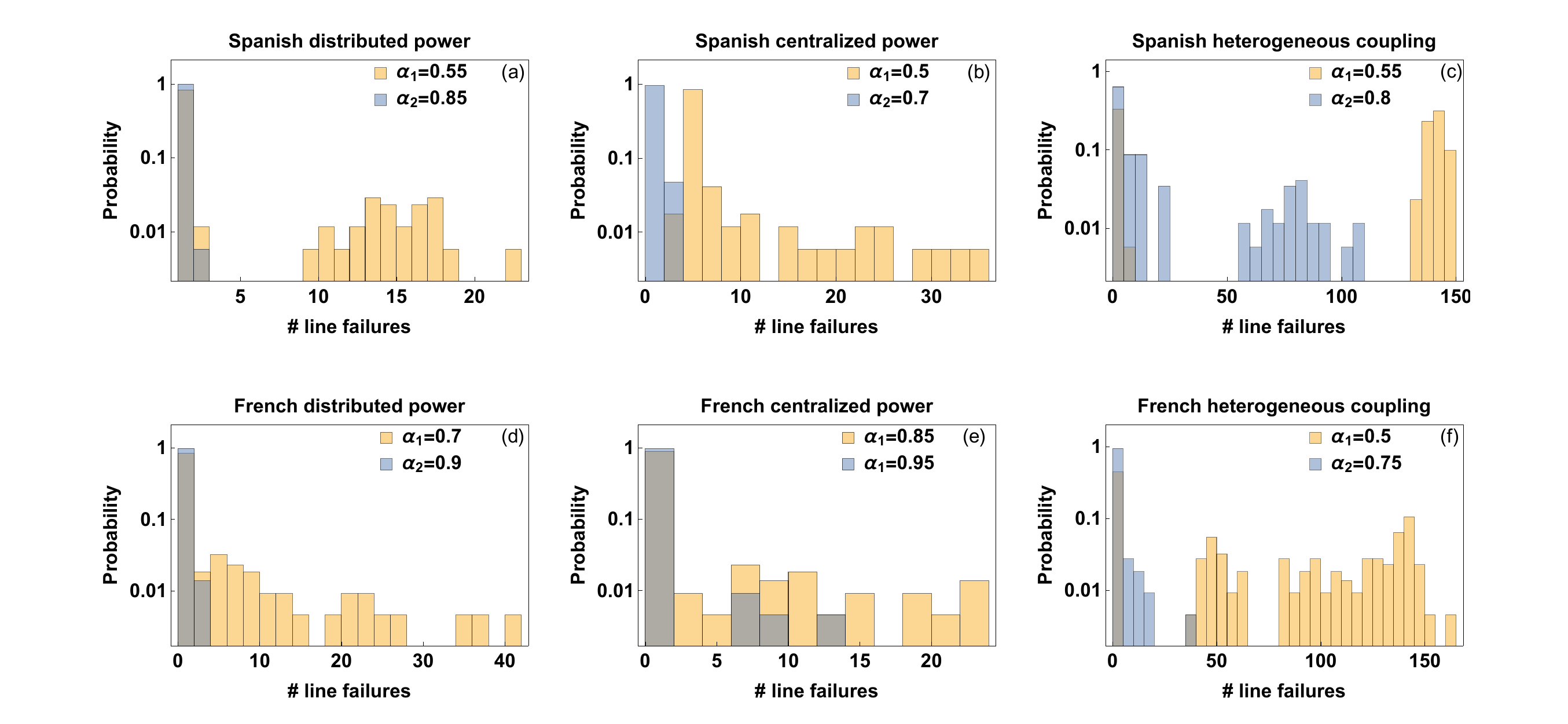}
\par\end{centering}
\caption{Network damage distributions in the Spanish and French power grids considering  different parameter settings. The histograms shown have been obtained
  under three different settings. Panels (a) and (d) refer to the case of  distributed power, 
  i.e., equal number of generators and consumers, each with $P^+=1/s^{2}$ and $P^-=-1/s^{2}$,
  and homogeneous coupling with $K=5/s^{2}$ for the Spanish and $K=8/s^{2}$ for the French grid.
  Panels (b) and (e) refer to the case of centralized power, i.e., consumers with $P^-=-1/s^{2}$
  and fewer but larger generators with $P^+\approx6/s^{2}$, and homogeneous coupling with $K=10/s^{2}$
  for Spanish and $K=9/s^{2}$ for the French grid. Panels (c) and (f) refer to a case of distributed power
  as in panel (a) and (d), but with  heterogeneous coupling, so
that the fixed point flows on the lines are approximately $F\approx0.5K$
both for the Spanish and the French grid. For all plots we use two different tolerances $\alpha$,
where the lower one is the smallest simulated value of $\alpha$ so that there are no initially
overloaded lines ($N-0$ stable).\label{fig:Histograms of line cuts}}
\end{figure*}

Furthermore, we have explored the role of centralized versus
distributed power generation, and that of heterogeneous couplings
$K_{ij}$, and also extended our analysis to other network topologies of European national power grids, namely those of
France and of Great Britain, see Supplementary Note 2.  In
Fig.~\ref{fig:Histograms of line cuts} we compare the results obtained
for the Spanish network topology (three top panels) to those obtained
for the French network (three bottom panels). With
$N_\text{French}=146$ nodes and $\left|E \right|_\text{French}=223$
edges the French power grid is larger in size than the Spanish one
considered in the previous figures ($N_\text{Spanish}=98$ and $\left|
E \right|_\text{Spanish}=175$) and has a smaller clustering
coefficient.  In each case, we have calculated the total number of line
failures at the end of the cascading failure when any possible line of the
network is used as the initial trigger of the cascade. We then plot
the probability of having a certain number of line failures in the
process, so that the histogram reported indicates the size of the
largest cascades and how often they occur. Notice that the probability
axis uses a log-scale.  For each network we have considered both
distributed and centralized locations of power generators, and both
homogeneous and heterogeneous network couplings. The centralized
generation is thereby a good approximation to the classical power grid
design with few large fossil and nuclear power plants powering the
whole grid. In contrast, the distributed generation scheme describes
well the case in which many small (wind, solar, biofuel, etc.)
generators are distributed across the grid
\cite{12powergrid}. Finally, the choice of heterogeneous coupling is
motivated by economic considerations, since maintaining a
transmission network costs money and only those lines that actually
carry flow are used in practice. In particular, we have worked
under the following three different types of settings:

First, we consider  distributed power and homogeneous coupling with an equal number of generators and consumers in the network, each of them having respectively
  $P^+=1/s^{2}$ and $P^-=-1/s^{2}$. The network uses homogeneous coupling with $K_{ij}=K a_{ij}$ and 
  $K=5/s^{2}$ for the Spanish (as in case of the previous figures) and $K=8/s^{2}$ for the French grid.
  Results for this case are shown in Fig.~\ref{fig:Histograms of line cuts} (a) and (d).
Next, we investigate centralized power and homogeneous couplings with consumers with $P^-=-1/s^{2}$ and fewer but larger generators
  with $P^+\approx6/s^{2}$.  The network uses homogeneous coupling with $K=10/s^{2}$
  for the Spanish and $K=9/s^{2}$ for the French grid.  Results for this case are shown
  in Fig.~\ref{fig:Histograms of line cuts} (b) and (e).
Finally, we apply distributed power and heterogeneous coupling with homogeneous
  distribution of generators and consumers as in case 1. The network uses a heterogeneous
 distribution of the $K_{ij}$, so that the fixed point
  flows on the lines are approximately $F\approx0.5K$ both for the
  Spanish and the French grid, see Supplementary Note 1 for details. 
  Results for this case are shown in Fig.~\ref{fig:Histograms of line cuts} (c) and (f).

In each of the above cases, we work in conditions such that no line is
overloaded before the initial exogenous damage.  We have performed
simulations for two values of the tolerance parameter $\alpha$. For
each of the two grids and of the three conditions above, the lowest
value $\alpha=\alpha_1$ has been selected to be equal to the minimal
tolerance such that each the network is $N-0$ stable (yellow
histograms). In addition, we have considered a second, larger value of
the tolerance, $\alpha_2$, showing qualitatively different behaviors
(blue histograms). As found in other studies
\cite{12powergrid,13powerlong,14bifurcation,16critical}, the
(homogeneous) coupling $K$ has to be larger for centralized generation
compared to distributed small generators to achieve comparable
stability. 

Initial line failures mostly do not cause any cascade and if they do, cascades typically affect only a small number of lines, see Fig.~\ref{fig:Histograms of line cuts}(a).  
This means that 
the Spanish grid is in most of the cases $N-1$ stable even in our dynamical model of cascades. Nevertheless, for $\alpha_1$,  
there exist a few lines that, when damaged,
trigger a substantial part of the network to be disconnected. 
This leads to the question whether and how the distribution of generators or the topology
of the network impact the size and frequency of the cascade.
When comparing distributed (many small generators) in panel (a) to centralized power
generation (few large generators) in panel (b) we do not observe a significant
difference in the statistics of the cascades. 
The same holds when comparing different network topologies, such as the Spanish and the French grid in
panels (d) and (e).

Conversely, allowing heterogeneous couplings introduces notable
differences to emerge in panels (c) and (f).  To obtain
heterogeneous couplings, we have scaled $K_{ij}$ at each line
proportional to the flow at the stable operational state, see
Supplementary Note 1. Thereby, we try to emulate cost-efficient grid
planning which only includes lines when they are used. However,
our results show that, under these conditions, the flow on a line with
large coupling cannot easily be re-routed in our heterogeneous network
when it fails \cite{16critical}.  For certain initially damaged lines,
this leads to very large cascades in grids with heterogeneous
coupling $K_{ij}$. For instance, both the Spanish and the French power
grid show a peak of probability corresponding to cascades of about 150
line failures when $\alpha=\alpha_1$. 
But also in the case of
$\alpha_2=0.8$, which corresponds to a $N-1$ stable situation under the homogeneous coupling condition, 
the Spanish grid exhibits cascades involving from 50 to 100 lines in $5 \%$ of the cases under heterogeneous couplings, see panel (c).  
The final number of
unsynchronized nodes after the cascade, used as a measure of the network damage follows qualitatively a similar statistics. 
Namely, distributed and centralized power generation return similar 
statistical distributions of damage, while under heterogeneous couplings the
system behaves differently. Furthermore, for each network, we have recorded
the two extreme situations in which either all nodes or the
grid stay synchronized, or the whole grid desynchronizes, see Supplementary Note 2. 

What do the results obtained here imply about the robust operation of power grids?
We have shown that
a network that is initially stable ($N-0$ stability), and remains
stable even to the initial damage of a line ($N-1$ stability)
according to the standard static analysis of cascades, can display  
large-scale dynamical cascades when properly modeled.
Although these dynamical overload events often have a very low
probability, their occurrence cannot be neglected since they
may collapse the entire power
transmission network with catastrophic consequences. 
In the examples studied, we have found that some critical lines cause
cascades resulting in a loss of up to $85\%$ of the edges (Fig.~\ref{fig:Histograms of line cuts}(c)). Hence, it is extremely important to develop methods to identify
such critical lines,
which is the subject of the next section.

\subsection*{Identifying critical lines}
\label{sec:CriticalLinks}

The statistical analysis presented in the previous section revealed that the size of the cascades triggered by different line
failures is very heterogeneous.  
Most lines of the networks investigated are not
critical, i.e., they are either $N-1$ stable even in our dynamical
model of cascades, or cause only a very small number of secondary
outages. However, for heavily loaded grids, as reported in Fig.~\ref{fig:Histograms of line cuts}, 
some highly critical lines emerge. Thereby, the initial failure of a single
transmission line causes a global cascade with the desynchronization
of the majority of nodes, leading to large blackouts. 
The key question here is whether
it is possible to devise a fast method to identify the critical lines of a 
network. This might prove to be very useful when it comes to improving the robustness of the network.  
\begin{figure}
\begin{centering}
\includegraphics[width=0.9\columnwidth]{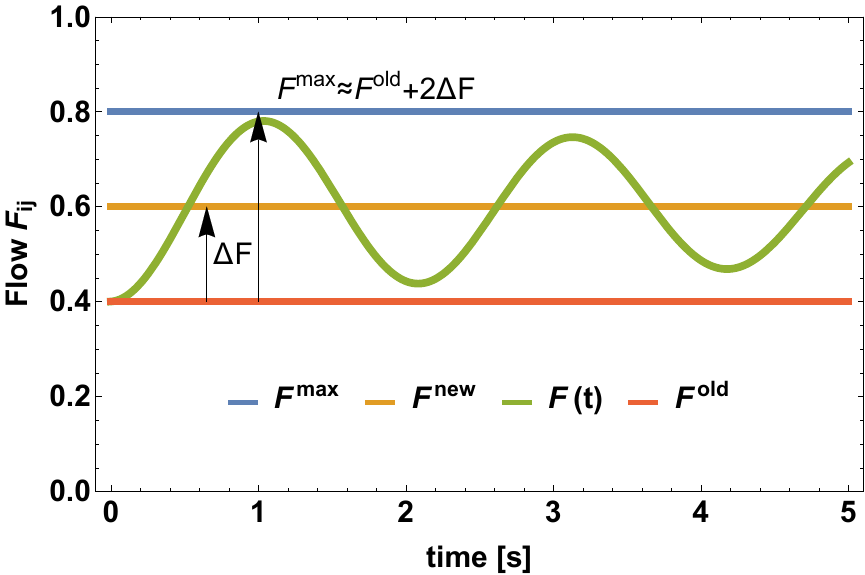}
\par\end{centering}
\caption{Introducing a flow-based estimator of the onset of a cascade. 
When cutting an initial line, the flows on a typical edge $(i,j)$ of the network 
increase from $F^{\text{ old}}$(red line) to $F^{\text{ new}}$ (orange line).
Based on numerical observations, the transient flow $F\left(t\right)$ from the old to the new
fixed point are well approximated as sinusoidal damped
oscillation. Knowing the fixed point
flows, allows to compute the difference $\Delta F =F^{\text{new}}-F^{\text{old}}$
and estimate the maximum
transient flow as $F^{\text{max}}\approx F^{\text{ old}}+2\Delta F$.
This estimation is typically slightly larger than the real flow because
the latter is damped. 
\label{fig:Illustration of cascade predictor}}
\end{figure}
In this section, we introduce a flow-based indicator for the onset of a cascade and
demonstrate the effectiveness of its predictions by comparing them to  results of the numerical simulation.
In particular, we show that our indicator is capable of identifying the critical links of the network
much better than other measures purely based on the topology or steady state of the
network, such as the edge betweenness \cite{Crucitti2004,Crucitti2004a,Hines2010}. 

\bigskip
In order to define a flow-based predictor for the onset of a cascading failure,
let us consider the typical time evolution of the flow along a line
after the initial removal of the first damaged line $(a,b)$.
As illustrated in Fig.~\ref{fig:Illustration of cascade predictor},
we observe flow oscillations after the initial line failure, which  are
well approximated by a damped sinusoidal function of time. See also
Supplementary Note 4 for the time evolution of the flows for the
case of the $N=5$ node graph introduced in Fig.~\ref{fig:Example of
  dynamical overload}.  Now, the steady flows of the network before
and after the removal of the trigger line are obtained by solving
Eq.~(\ref{eq:stable state}) for the fixed point angles $\{
\theta_{i}^{*} \}$, which depend on the node powers $\{ P_{i} \}$ and
on the coupling matrix $\{ K_{ij} \}$. Thereby, we obtain a set of
nonlinear algebraic equations which have at least one solution if the
coupling $K$ is larger than the critical coupling
\cite{14bifurcation}. 
For sufficiently large values of the coupling $K$ there can be 
multiple fixed points \cite{Manik2016a}. In each case, we
determine a single fixed point with small initial flows by using
Newton's method, see Supplementary Note 1 for details.
From the values of the fixed point angles $\{ \theta^{*} \}$ we
calculate the equilibrium flow along each line, for instance line
$(i,j)$, before and after the removal of the trigger line, from the
expression:   
\begin{equation}
  F_{ij}^{\text{*}}=K\sin\left(\theta_{j}^{*}-\theta_{i}^{*}\right).
\end{equation}
Let us indicate the 
initial flow along line $(i,j)$ in the intact network as $F_{ij}^{\text{old}}$, and the new flow after the removal of the 
trigger line as $F_{ij}^{\text{new}}$, assuming there still is a fixed point. 
Given enough time, the system settles in the new fixed
point and the change of flow on the line is $\Delta F_{ij}=F_{ij}^{\text{new}}-F_{ij}^{\text{old}}$.
Based on the oscillatory behavior observed in cascading events, see Fig.~\ref{fig:Illustration of cascade predictor} for an illustration,
we approximate the time-dependent flow on the line close to the new fixed point as:  
\begin{equation}
F_{ij}\left(t\right)\approx F_{ij}^{\text{new}}-\Delta F_{ij}\cos\left(\nu_{ij}t\right)e^{-D t},
\end{equation}
where $\nu_{ij}$ is the oscillation frequency specific to the link $(i,j)$ and
$D$ is a damping factor. 
The maximum flow  $F_{ij}^{\text{max}}$ on the line during the transient
phase is then given by:  
\begin{equation}
 F_{ij}^{\text{max}} \approx F_{ij}^{\text{old}}+2\Delta F_{ij}.\label{eq:Analytical max flow}
\end{equation}
Hence, for the cascade predictor we propose to test whether a line will be overloaded during the
transient by computing $F_{ij}^{\text{max}}$ from the expression above and by 
checking whether $F_{ij}^{\text{max}}$ is larger than the available capacity $C_{ij}$ of the link. 
This procedure provides a good approximation of the real flows. However, it requires fixed point calculations of the intact network and of the network after the initial trigger line is removed.
Furthermore, it has to be repeated for each possible initial trigger
line, so that a total of $\left|E\right|+1$ fixed points is being computed, 
with $\left|E\right|$ being the number of edges. A possible way
to simplify this procedure is to compute the fixed point flows of
the intact grid $F_{ij}^{\text{old}}$ only,   
approximating the fixed point flows after changes of the network topology
by the Line Outage Distribution Factor (LODF) \cite{Manik2017,Ronellenfitsch2017}.
Details on this method can be found in Supplementary Note 1.

After starting the cascade by removing line $(a,b)$, we define our analytical prediction
for the minimal transient tolerance $\left(\alpha^{\text{tr.}\,(a,b)}_{ij}\right)_\text{min}$ based on the maximum transient flow on line $(i,j)$ given in
Eq.~(\ref{eq:Analytical max flow}): 
\begin{equation}
\left(\alpha^{\text{tr.}\,(a,b)}_{ij}\right)_\text{min} = F_{ij}^{\text{max}}\label{eq:analytical minimal tolerance}
\end{equation}
such that, if $\alpha > \left(\alpha^{\text{tr.}\,(a,b)}_{ij}\right)_\text{min}$, then cutting line $(a,b)$ as a trigger  will
not affect line $(i,j)$. Finally, 
we define the minimal tolerance $\left(\alpha^{\text{tr.}\,(a,b)}\right)_\text{min}$ of the network as that
value of $\alpha$ such that there is no secondary failure after the initial failure of
the trigger line, i.e., the grid is $N-1$ secure. We have: 
\begin{equation}
 \left(\alpha^{\text{tr.}\,(a,b)}\right)_\text{min} =  \max_{(i,j)} \left(\alpha_{ij}^{\text{tr.}\,(a,b)}\right)_\text{min}
                    =  \max_{(i,j)} \left(   F_{ij}^{\text{max}}    \right), 
\label{eq:network minimal tolerance}
\end{equation}
where the maximum is taken with respect to all links $(i,j)$ in the
network and one trigger link $(a,b)$. If we set $\alpha \ge \left(\alpha^{\text{tr.}\,(a,b)}\right)_\text{min}$ then, according
to our prediction method, we expect no additional line failures further to
the initial damaged line.  Let us assume that the network topology is
given, for instance that of a real national power grid, and that the
tolerance level is preset due to external constrains like security
regulations. Then, the calculation of
$\left(\alpha^{\text{tr.}\,(a,b)}\right)_\text{min}$ allows to engineer a resilient 
grid by trying out different realizations of $K_{ij}$. When changes of
$K_{ij}$ are small, the new fixed point flows are approximated by
linear response of the old flows \cite{Manik2017} giving us an easy
way to design the power grid to fulfill safety requirements.

\bigskip

\begin{figure*}
\begin{centering}
\includegraphics[width=2.1\columnwidth]{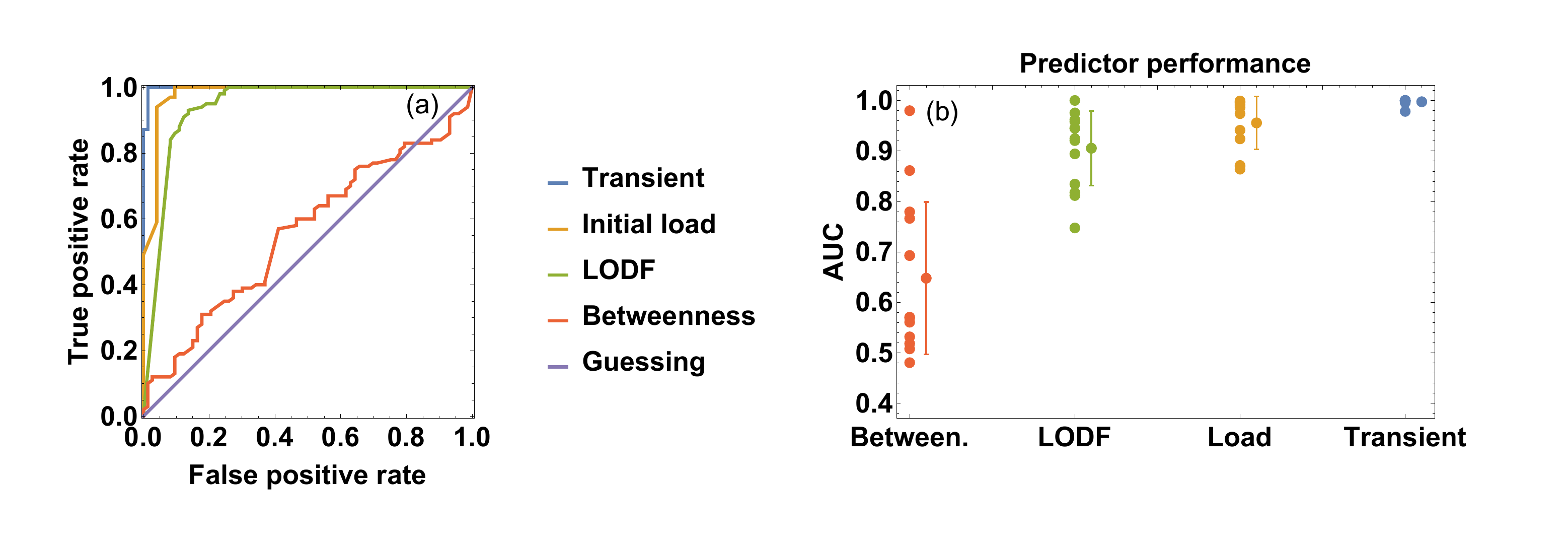}
\par\end{centering}
\caption{Comparing the predictions of the flow-based indicator of critical lines to
  other standard measures. Four different predictors are presented to determine whether a
given line, if chosen as initially damaged, causes at least one additional line failures. 
Our dynamical predictor (indicated as Transient) is based on the estimated maximum transient
flow (\ref{eq:Analytical max flow}). The predictor based on the Line Outage Distribution
Factor (LODF) \cite{Manik2017,Ronellenfitsch2017} uses the same idea
but computes the new fixed flows based on a linearization of the flow computation.
Predictors based on betweenness and initial load classify a line as critical if it is within
the top $\sigma^\text{thr}\times 100\%$ of the edges with highest betweenness/load
 with threshold $\sigma^\text{thr}\in\left[0,1\right]$.
Panel (a) shows the ROC curves obtained for the Spanish grid with heterogeneous
coupling and tolerance $\alpha=0.7$, while in panel (b) the 
AUC is displayed for 
all network settings presented in Fig.~\ref{fig:Histograms of line cuts}.  For each predictor all individual scores are displayed on the left and the mean with error bars based
on one standard deviation is shown on the right. 
\label{fig:ROC curve Spanish 5 analytics} }
\end{figure*}

To measure the quality of our predictor for critical lines and to compare it 
to alternative predictors, we quantify its performance by
evaluating how often it detects critical lines as critical (true positives) compared
to how often it gives false alarms (false positives). In our model for cascading
failures, a potential trigger line is classified as truly critical if its removal
causes additional secondary failures in the network according to the numerical
simulations of the dynamics \cite{16critical}. The flow-based prediction is
obtained by first calculating the minimal tolerance of the network
$\left(\alpha^{\text{tr.}\,(a,b)}\right)_\text{min}$
based on Eq.~(\ref{eq:network minimal tolerance})
 and comparing it with the fixed tolerance $\alpha$ of
 a given simulation. If the obtained minimal tolerance is larger
 than the value of tolerance used in the numerical simulation, than 
 the line is classified as critical by our predictor 
 and additional overloads are to be expected.
More formally, we use the following prediction rules: 
\begin{eqnarray}
  \left(\alpha^{\text{tr.}\,(a,b)}\right)_\text{min} & \geq & \alpha + \sigma^\text{thr} \,\Rightarrow\text{critical},
  \\
\left(\alpha^{\text{tr.}\,(a,b)}\right)_\text{min} & < & \alpha + \sigma^\text{thr}\,\Rightarrow\text{not critical,}
\label{eq_critical}
\end{eqnarray}
with a variable threshold  $\sigma^\text{thr}$   $\in\left[-1,1\right]$, which allows to tune the sensitiviy of the predictor.

\medskip
Analogously, we define a second predictor based on the Line Outage
Distribution Factor (LODF) \cite{Ronellenfitsch2017,Manik2017}. In this case, the expected minimal tolerance is obtained by approximating
the new flow by the LODF, instead of computing them
by solving for the new fixed points, see Supplementary Note 1.

\medskip
We compare our predictors based on the flow dynamics to the 
pure topological (or steady-state based) measures that have been used
in the classical analysis of cascades on networks. The idea
behind such measures is the following.  
First, we consider the initial load on all potential trigger lines $(a,b)$: $L^{(a,b)}=F_{ab}(t=0)$, i.e., the flow at time $t=0$ on the line,
when the system is in its steady state. 
Intuitively, highly loaded lines are expected to be more
critical than less loaded ones. Hence, comparing each load $L^{(a,b)}$ to
the maximum load on any line in the grid $L_\text{max}:=\max_{(i,j)}L^{(i,j)}$ leads to the following
prediction: 
\begin{eqnarray}
L^{(a,b)} & \geq & \left(1- \sigma^\text{thr} \right)L_\text{max}\,\Rightarrow\text{critical,}
\\
L^{(a,b)} & < & \left(1- \sigma^\text{thr} \right)L_\text{max}\,\Rightarrow\text{not critical},
\end{eqnarray}
where $\sigma^\text{thr}\in\left[0,1\right]$ is the prediction threshold.

Another quantity that is often used as a measure of the importance of a network edge
is the edge betweenness \cite{Newman2010,Latora2017}. The betweenness  $b^{(a,b)}$ 
of edge $(a,b)$ is defined as the normalized number of shortest paths passing by the edge. 
A predictor based on the edge betweenness $b^{(a,b)}$ is then obtained by replacing
$L^{(a,b)}$ by $b^{(a,b)}$ in the expressions above.

\medskip
To evaluate the predictive power of the flow-based cascade predictors and to compare
them to the standard topological predictors, we have computed the number of lines that
cause a cascade by simulation and compared how often each predictor correctly predicted
the cascade, thereby deriving the rate of correct cascade predictions (true positive
rate) and rate of false alarms (false positive rate). 
These two quantities are displayed in a Receiver Operator Characteristics
(ROC) curve, which reports the true positive rate versus the false positive
rate when varying the threshold $\sigma^\text{thr}$. The ROC curve would go up straight from point $(0,0)$ to point $(0,1)$ in the
ideal case in which the predictor is capable of detecting all real cascade events,  
while never giving a false positive. 
Conversely, random guessing corresponds to
the bisector.
Finally, any realistic predictor starts at the point $(0,0)$, i.e. never
giving an alarm regardless of the setting, and evolves to the point
$(1,1)$, i.e. always giving an alarm.
The transition from $(0,0)$ to $(1,1)$ is tuned by decreasing the threshold
$\sigma^\text{thr}$ determining when to give an alarm.

The ROC curves corresponding to the predictors introduced above are
shown in Fig.~\ref{fig:ROC curve Spanish 5 analytics} (a). A  
prediction based on the betweenness of the line is only
as good as a random guess. In contrast, using the LODF
and the initial load provide much better 
predictions. Finally, the analytical prediction outperforms any
other method, well approximating an ideal predictor.
\begin{figure*}[t]
\begin{centering}
\includegraphics[width=2\columnwidth]{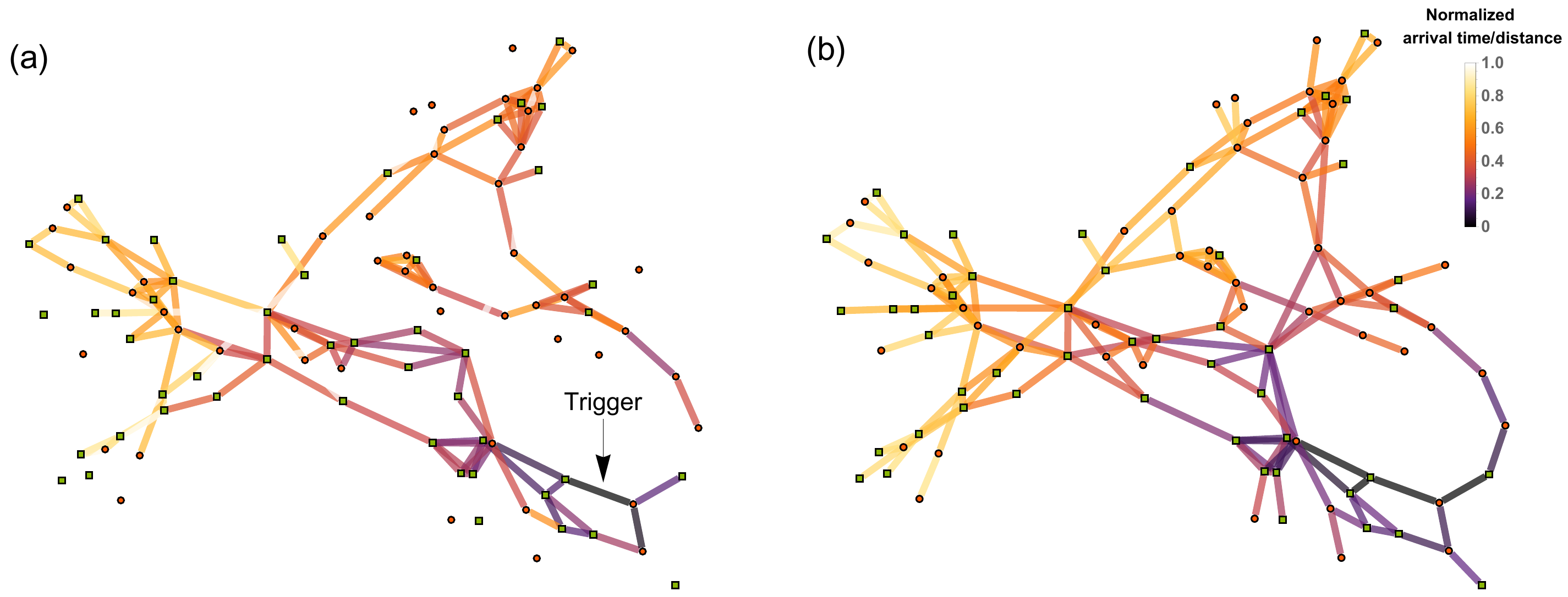} \par\end{centering}
\caption{Mapping the propagation of a cascade on the Spanish power grid.  
(a) The edges of the network are color-coded 
based on the normalized arrival time
of the cascade with respect to a specific initially damaged
line, indicated as ``Trigger''., and  
(b) based on their normalized distance with respect to the trigger using the effective distance measure in 
Eq.~(\ref{eq:Effective distance definition}).  In both cases, darker colors indicate shorter distance/early arrival
of the cascade. Normalization is carried out using the largest distance/arrival
time. Edges that are not plotted are not reached by the cascade at
all. The analysis has been performed using the Spanish grid with distributed generators with $P^+=1$
(green squares), consumers with $P^-=-1/s^{2}$ (red circles), 
heterogeneous coupling 
and tolerance $\alpha=0.55$. \label{fig:Cascade propagation color map} }
\end{figure*}

An alternative way to quantify the quality of a predictor is by
evaluating the Area Under Curve (AUC), that is the size of
the area under the ROC curve. An ideal predictor would correspond to
the maximum possible value AUC$=1$, while a random guess produces an
AUC of $0.5$.  So the closer the value of AUC for a given predictor is
to 1, the better are the obtained predictions. AUC scores have
been computed for different networks, settings and parameters.  The
results for the dynamical flow-based predictor, the predictor based on
the LODF, as well as the initial load and betweenness predictors,
are shown in Fig.~\ref{fig:ROC curve Spanish 5 analytics} (b). The 
values of the AUC scores reported correspond to the different settings
described in Fig.~\ref{fig:Histograms of line cuts}, allowing a more
systematic comparison of predictors than that provided by a single ROC
curve. Also from this figure it is clear that a prediction of the
critical links based on their betweenness is on average only slightly
better than random guessing. Furthermore, this result rises concerns on the
indiscriminate use of the betweenness as a measure of centrality in
complex networks. Especially when the dynamical processes of interest
are well known, this must be taken into account in the definition of
dynamical centrality measures for complex networks \cite{12klemm,Hines2010,Hines2008}.
The LODF and initial load predictors perform relatively better on average,
although they still display  large standard deviations. 
This means that,
for certain networks and settings they reach an AUC score close to the
perfect value of $1$, while in some other cases they only reach values
of AUC equal to $0.8$. Of these two indicators, the initial load
predictor results are more reliable. Finally, our dynamical predictor, indicated in figure as ``Transient''
outperforms all alternative ones, in every single parameter and
network realization. The figure indicates that the corresponding AUC scores
reach values very close to $1$. Moreover, this indicator displays the
smallest standard deviation when different networks and parameter settings are
considered. In conclusion, this seems to be the best indicator for the
criticality of a link. However, the results show that, although the initial
load predictor performs worse than our dynamical one, it might still be used when
computational resources are scarce as it provides the second best
predictions among those considered.

\subsection*{Cascade propagation}
\label{sec:CascadePropagation}

So far, we have shown that network cascades, i.e., secondary failures 
following an initial trigger, can well be caused by transient dynamical effects. 
We have proposed a model for power grids that takes this into account, and  we have also developed a
reliable method to predict whether additional lines can be affected by
an initial damage, potentially triggering a cascade of failures.
However, knowing whether a cascade develops or not does not answer
another important question that is to understand how the cascade
evolves throughout the network, and which nodes and links are affected and
when. Intuitively, we expect that network components farther away from
the initial failure should be affected later by the cascade. 
We have indeed observed that the time a line fails and its distance from the initial triggering link are correlated. Instead of merely using the graph topology to measure distances, we use a more sophisticated distance measure, the effective distance, based on the characteristic 
flow from one node to its neighbors.
This idea has been first 
introduced in Ref.~\cite{Brockmann2013} in the context of disease
spreading, where the effective distance has been shown
to be capable of capturing spreading phenomena better than the standard graph
distance. The effective distance between two 
vertices $i$ and $j$ can be defined in our case as: 
\begin{equation}
 d_{ij}=1-\log\left(\frac{K_{ij}}{\sum_{k=1}^{N}K_{ik}}\right).
\label{eq:Effective distance definition}
\end{equation}
Here, we used the coupling matrix $K_{ij}$ as a measure of the flows between nodes \cite{Brockmann2013}.
All pairs of nodes not sharing an edge, i.e. such that $K_{ij}=0$, have infinite
effective distance $d_{ij}=\infty$. At each node the cascade spreads
to all neighbors but those that are coupled tightly, get affected the
most and hence get assigned the smallest distance $d_{ij}$. Furthermore,
the effective distance is an asymmetric measure, since $d_{ij}\neq
d_{ji}$ in general. The quantity $d_{ij}$ is a property of two
nodes, while the most elementary damage in our cascade model 
affects edges. Hence, the concept of distance has to be extended 
from couples of nodes to couples of links. For instance, in the case
of an unweighted network it is possible to define the (standard) distance
between two edges as the number of hops along a shortest path connecting
the two edges. In the case of a weighted graph we make use of the measure
of effective distance in Eq.~(\ref{eq:Effective distance definition}) to
define a distance between two edges as the minimal path length
of all weighted shortest paths between two edges.
The distance between two edges can then be obtained based on the
definition of distances between nodes $\{ d_{ij} \}$.
Given the trigger edge $(a,b)$, the distance from edge $(a,b)$ to edge $(i,j)$ 
is given by:
\begin{equation}
d_{(a,b)\rightarrow(i,j)}= d_{ab} + \min_{v_{1}\in\left\{ a,b\right\} ,v_{2}\in\left\{ i,j\right\} }d_{v_{1}v_{2}} \label{eq:edge distance}
\end{equation}
i.e., it is the minimum of the shortest path lengths of the paths $a\rightarrow i$,
$a\rightarrow j$, $b\rightarrow i$ and $b\rightarrow j$, plus the
effective distance between the two vertices $a$ and $b$.
\begin{figure}
\begin{centering}
\includegraphics[width=0.85\columnwidth]{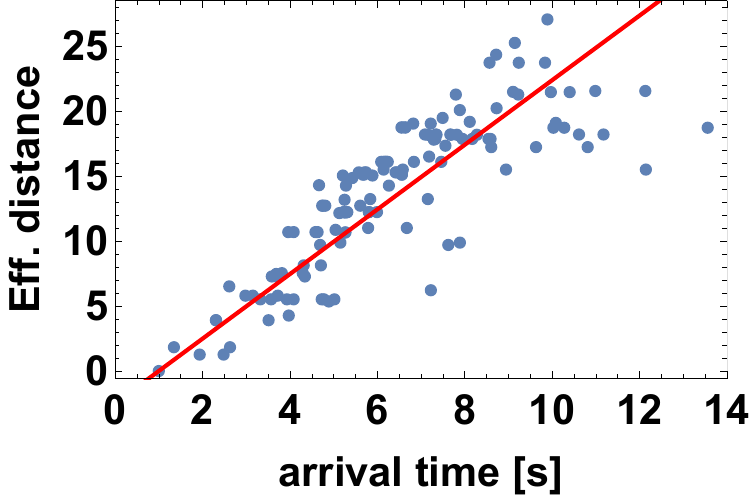}
\par\end{centering}
\caption{
  Effective distances between the initial trigger and 
  secondary line outages are plotted as a function of time. Each point in the
  plot represents one edge, while the straight line is the result of
  a linear fit.  The reported fit indicates that
the two quantities are related by an approximate linear relationship
with regression coefficient $R^2\approx 0.94$. 
  Results refer to the Spanish power grid with the same
  parameters and trigger as used in Fig.~\ref{fig:Cascade
    propagation color map}.
\label{fig:Cascade propagation plot}}
\end{figure}

Fig.~\ref{fig:Cascade propagation color map} shows that the effective
distance is capable of capturing well the properties of the spatial
propagation of the cascade over the network from the location of the
initial shock. The figure refers to the case of the Spanish grid topology
with heterogeneous coupling (see Figs. \ref{fig:Spanish map and single line cuts} and \ref{fig:Histograms of line cuts}). The temporal evolution of one
particular cascade event, which is started by an initial exogenous
damage of the edge marked as ``Trigger'', is reported.   Network edges are color-coded based on the
actual arrival time of the cascade in panel (a), and compared to a
color code based instead on their effective distance from the trigger
line in panel (b).  Edges far away from the trigger line, in terms of
effective distance, have brighter colors than edges close to the
trigger. Similarly, lines at which the cascade arrives later are
brighter than lines affected immediately. The figure clearly indicates
that effective distance and arrival time are highly correlated, i.e.,
the cascade propagates throughout the network reaching earlier those
edges that are closer according to the definition of effective
distance. The relation between the effective distance of a line from
the initial trigger and the time it takes for this line to be affected
by the cascade is further investigated in the scatter plot of
Fig.~\ref{fig:Cascade propagation plot}. 
We observe a substantial correlation between arrival time and shortest distance, indicating the possibility of an effective speed of cascading failures across the network. The mean correlation between cascade arrival time and effective distance, $\left< R_\text{eff.}^2 \right>\approx0.91$, is
larger than between the arrival time and simple graph-theoretic distance, $\left< R_\text{graph}^2 \right>\approx0.88$. See Supplementary Note 5 for details and \cite{Hines2017} for further discussion on propagation of cascades.


\section*{Discussion}
In this work, we have proposed and studied a model of electrical transmission networks
highlighting the importance of transient dynamical behavior in
the emergence and evolution of cascades of failures.
The model takes
into account the intrinsic dynamical nature of the system, in contrast
to most  other studies on supply networks, which
are instead based on a static flow analysis. 
Differently from the existing works on cascading failures in power
grids
\cite{Simonsen2008,Witthaut2015,Plietzsch2016,Rohden2016,Crucitti2004,Crucitti2004a,Kinney2005,Ji2016,Pahwa2014}, 
we have exploited the dynamic nature of the swing equation to
describe the temporal behavior of the system, and we  
have adopted an absolute flow threshold to model the propagation of a cascade
and to identify the critical lines of a network.  
The differences with respects to the results of a 
static flow analysis are striking, as $N-1$ secure power grids, i.e.,
grids for which the static analysis does not predict any additional
failures, can display large dynamical cascades. 
This result emphasizes the importance of taking dynamical transients of the order of seconds into
account when analyzing cascades, and should be considered by grid operators
when performing a power dispatch, or during grid extensions.
Notably, our dynamical model for cascades not only reveals additional
failures, but also allows to study the details of the spreading of
the cascade over the network. We have investigated such a propagation 
by using an effective distance measure quantifying the distance of a line (link of the network)
from the original failure, which strongly correlates with the time it takes for the cascade
to reach this line. 
The observed correlation between propagation time and effective
distance of a failure, points to the possibility of extracting an effective
speed of the cascade propagation. This result may thus stimulate further
research understanding propagation patterns on networks. Being able to measure the speed
of a cascade would further contribute to the design of measures to
stop or contain cascades in real time because such propagation speed
determines how fast actions have to be taken.
We remark that an approximately constant speed in terms of e.g. the effective distance measure, may represent a highly non-local spreading in terms of geographical distances, 
also observed in \cite{Hines2017,Yang2017a}.
Moreover, propagation patterns of line failures caused by current overloads, as investigated here, may be qualitatively different from those caused by voltage effects \cite{USBlackout2003,Simpson-Porco2016}. On longer time scales the operation of control systems and emergency measures such as load shedding must be taken into account to assess the impact of a cascade of failures. These features are typically studied in quasi-static models such that the short time scale considered in this paper offers a complementary view to the  spreading of cascading failures.

While the swing equation is capable of capturing interesting dynamical effects previously
unnoticed, it still constitutes a comparably simple model to describe 
power grids \cite{Mach08}. Alternative, more elaborated models would involve
more variables, e.g., voltages at each node of the network to allow a description of longer time scales \cite{Auer2016,Schmietendorf2014,Sharafutdinov2017,Ma2016}.
In addition,  
we only focused on the removal of individual lines in our framework, 
instead of including the shutdown of power plants, i.e., the
removal of network nodes. These simplifications are mainly justified by the
very same time scale of the dynamical phenomena. Most 
cascades observed in the simulation are very fast, terminating on a time scale of less than $10$ seconds, which
supports the choice of the swing equation
\cite{Mach08,Kund94}. Furthermore, such short time scales are consistent
with empirical observations of real cascades in power grids, 
which were caused in a very short time by overloaded lines. Conversely,  
power plants (nodes of the network) were usually shut 
down after the failure of a large fraction of the transmission grid. 
The same holds for load shedding, i.e., disconnecting consumers. 
Summing up, while the overall blackout takes place over minutes, critical 
damage is done within seconds due to line failures  \cite{Bialek2007,USBlackout2003,UCTE_Report2007}. Hence, this article models the short time scale of line failures only.

In order to further
support our conclusions, we have considered additional models and discussed the validity of the swing equation in Supplementary Note 3. 
In particular, we have also simulated a third order model that includes voltage dynamics, finding qualitatively similar 
results to those obtained with the swing equation.
Furthermore, a recent study \cite{Cetinay2017} also highlights that a DC approach misses important events, 
and an AC model is necessary to capture all aspects of cascades. While the authors in \cite{Cetinay2017} use realistic (IEEE) grids and more detailed simulation models, we complement this numerical approach by providing semi-analytical insight into cascades. Specifically, we provide simple predictors of critical lines and observe a propagating cascade.
Overall, our work indicates 
that a dynamical second order model, as the one adopted in our framework, is capable of
capturing additional features compared to static flow analyses,
while still making analytical approaches possible. This allows to 
go beyond the methods commonly adopted in the engineering literature,
which are often solely based on heavy computer simulations of specific
scenarios, e.g. \cite{Salmeron2004}.

Furthermore, concerning the delicate issue of protecting the grid
against random failures or targeted attacks, it is crucial to be able
to identify critical lines whose removal might be causing large-scale
outages. As we have seen, most of the lines of the networks studied in
this article cause very small cascades when initially damaged.
However, our results have also unveiled the existence in each of these
networks of a few critical lines producing large outages, which in
certain cases can even affect the entire grid. Within our modeling
framework, we have been able to develop an analytical
flow predictor that reliably identifies critical lines and
outperforms existing topological measures in terms of prediction
power. As an alternative to the  analytical flow predictor, when a  
faster assessment of criticality is required, the stable state flows
of the intact grid can be used, although they are less reliable.
We hope these two indicators can become a useful tool for grid
operators to test their current power dispatch strategies against
cascading threads.

In a time when our lives depend more than ever on the proper functioning 
of supply networks, we believe it is crucial to understand their
vulnerabilities and design them to be as robust as possible.
The results presented in this article represent only a first step
in this direction and many interesting questions  
remain to be investigated and answered within our framework or
similar approaches. 
If cascades often propagate non-locally, can a quantity like a propagation speed  be defined or is it otherwise possible to  predict cascade arrival times in a unified way?
Which lines are affected by a large cascade, and which parts of the
network are capable of returning to a
stable state? What are the best mitigation strategies to contain a
cascade or to stop its propagation?
All these questions go beyond the scope of this article, whose
aim was mainly to provide a first broad analysis of the importance 
of transients in the emergence and evolution of cascades, but we
hope our results will trigger the interest of the research
community of physicists, mathematicians and engineers.   

\section*{Methods}

\subsection*{Modeling Power Grids}
When it comes to model the dynamics of a power transmission
  network, the swing equation is a simple way to
  deal with the key features of the system as a whole, namely its dynamical 
  synchronization properties. Thereby, we avoid dealing directly with a
  complete dynamical description in terms of complicated
  power grid simulation software or static power-flow models which are
  routinely used to simulate specific scenarios on large-scale power grids by power
  engineers.
The swing equation retains the dynamical features of AC power grids, 
by describing each of the elements of an electric power network as 
a rotating machine characterized by its angle  and its
angular velocity at a given time. 
In practice, a rotating machine either represents a large synchronous generator in a conventional 
power plant or a coherent subgroup, i.e., a group of strongly coupled small machines and loads which 
are tightly phase-locked in all cases. Note that this is a coarse-grained model where every node is modeled as a rotating machine with effective inertia. A node with higher demand than supply will
then act as an effective consumer, i.e., a synchronous motor.
The angle of each machine is assumed to be identical to the angle
of the complex voltage vector, so that the angle difference of two
machines determines the power flow between them to transport,
for example, energy from a generator to a consumer.

More formally, let us suppose to have $N$ rotatory machines, each
corresponding to a node of a network.  Each machine $i$, with
$i=1,2,\ldots,N$ is characterized by its mechanical rotor angle
  $\theta_{i}(t)$ and by its angular velocity
  $\omega_{i}:=\text{d}\theta_{i}/\text{d}t$  
  relative to the
  reference frame of $\Omega=2\pi \left(50\mbox{ or }60\right)$
  Hz.
Furthermore, machine $i$ either feeds 
power into the network, acting as an effective generator with power
$P_{i}>0$, or absorbs power, acting as an effective consumer
(corresponding to the aggregate consumers of an urban areas) with power 
$P_{i}<0$. The  swing equation reads
\cite{Fila08,Mach08,Nish15}: 
\begin{eqnarray}
\frac{\text{d}}{\text{d}t}\theta_{i} & = & \omega_{i}\label{eq:Swing equation}\\
I_i\frac{\text{d}}{\text{d}t}\omega_{i} & = & P_{i}-\gamma_i\omega_{i}+\sum_{j=1}^{N}K_{ij}\sin\left(\theta_{j}-\theta_{i}\right),
\end{eqnarray}
where $\gamma_i$ is the homogeneous damping of an oscillator, $I_i$ is the inertia constant and
$K_{ij}$ is a coupling matrix governing the topology of the power grid
network, and the strength of the interactions. In the following, we
will both consider heterogeneous coupling $K_{ij}$ or we will assume 
homogeneous coupling $K_{ij}=K a_{ij}$, where $a_{ij}$ are the
entries of the unweighted adjacency matrix that describes the
connectivity of the network. 
For simplicity, we assume homogeneous damping $\gamma_i=\gamma$ and inertia $I_i=1$ for all $i\in {1,...,N}$.
To derive Eq.~(\ref{eq:Swing equation}) one has to assume 
that the voltage amplitude $V_{i}$ at each nodes is time-independent,
that ohmic losses are negligible and that the changes in the angular
velocity are small compared to the reference $\omega_{i}\ll\Omega$, see e.g, \cite{Fila08,Kund94} for details.
All these assumptions are fulfilled as long as we model short time
scales on the high-voltage transmission grid \cite{Mach08} which
will be sufficient for our study. The coupling matrix $K_{ij}$ is
an abbreviation for $K_{ij}=B_{ij}V_{i}V_{j}$ where $B_{ij}$ is
the susceptance between two nodes \cite{Kund94}. 
The swing equation is especially
well suited to describe short time scales, as they appear in typical
large-scale power grid cascades \cite{Bialek2007,USBlackout2003,UCTE_Report2007},
however, we also discuss other models returning qualitatively similar
results in Supplementary Note 3.

The desired stable state of operation of the power grid network 
is characterized by all machines running in synchrony at the reference
angular velocity $\Omega$, i.e., $\omega_{i}=0$ $\forall i\in\left\{ 1,...,N\right\} $, 
implying $\sum_i P_i=0$.
Thereby, we determine the fixed point by solving for the angles
$\theta_{i}^{*}$ in:  
\begin{equation}
0=P_{i}+\sum_{j=1}^{N}K_{ij}\sin\left(\theta_{j}^{*}-\theta_{i}^{*}\right).\label{eq:stable state}
\end{equation}

The grid in its synchronous state is phase-locked, i.e., all angle
differences do not change in time. This is important since the angle
difference determines the flow along a line, and fluctuating angle
differences would imply fluctuating conducted power which can in turn
lead to the shutdown of a plant \cite{Mach08,Kund94}. Furthermore, transmission
system operators demand the frequency to stay within strict boundaries
to ensure stability and constant phase locking \cite{entsoe14}.

Phase-locking and other synchronization phenomena arise in many
different domains and applications, and have attracted the interest of
physicists across fields \cite{Pikovsky2003}. One of the simplest
synchronization models is the Kuramoto model
which has been used, among other applications, to describe synchronization
phenomena in fireflies, chemical reactions and simple
neuronal models \cite{Kura75,Stro00,Witthaut2017}. 
The swing equation shows similarities with the Kuromoto model, including
the sinusoidal form of the coupling function and the existence of 
a minimal coupling threshold to achieve synchronization
\cite{14bifurcation}. However, the swing equation includes a second
derivative due to the inertial forces in the grid. 
Both equations share the same fixed points 
but the swing equations display dissipative forces and limit cycles that 
are not present in the Kuramoto model. 

\subsection*{Data availability}
The networks used in this study and examples of elementary cascades are provided at https://osf.io/jz4m6/. 
All data that support the results presented in the figures of this study are available from the authors upon request.

\begin{acknowledgments}
We thank Vittorio Rosato for providing the national grid topologies. 
B.S. and V. L. acknowledge support from the EPSRC project EP/N013492/1,  
``Nash equilibria for load balancing in networked power systems''. 
We gratefully acknowledge support from the Federal Ministry of Education
and Research (BMBF grant no.03SF0472A-F), the Helmholtz Association
(via the joint initiative ``Energy System 2050 - A Contribution of
the Research Field Energy'' and the grant no.VH-NG-1025 to D.W.),
the Göttingen Graduate School for Neurosciences and Molecular Biosciences
(DFG Grant GSC 226/2 to B.S.) and the Max Planck Society (to M.T.). 
Finally, we gratefully acknowledge support from the German Science Foundation (DFG) by a grant toward the Cluster of Excellence "Center for Advancing Electronics Dresden" (cfaed)
\end{acknowledgments}

\subsection*{Author contributions}
B.S. and V.L. designed the research. B.S. performed most simulations and generated figures. D.W. and M.T. contributed to discussing intermediate and final results and writing the paper. 

\subsection*{Competing interests}
The authors declare no competing financial or non-financial interests.


\begin{thebibliography}{0}%
\makeatletter
\providecommand \@ifxundefined [1]{%
 \@ifx{#1\undefined}
}%
\providecommand \@ifnum [1]{%
 \ifnum #1\expandafter \@firstoftwo
 \else \expandafter \@secondoftwo
 \fi
}%
\providecommand \@ifx [1]{%
 \ifx #1\expandafter \@firstoftwo
 \else \expandafter \@secondoftwo
 \fi
}%
\providecommand \natexlab [1]{#1}%
\providecommand \enquote  [1]{``#1''}%
\providecommand \bibnamefont  [1]{#1}%
\providecommand \bibfnamefont [1]{#1}%
\providecommand \citenamefont [1]{#1}%
\providecommand \href@noop [0]{\@secondoftwo}%
\providecommand \href [0]{\begingroup \@sanitize@url \@href}%
\providecommand \@href[1]{\@@startlink{#1}\@@href}%
\providecommand \@@href[1]{\endgroup#1\@@endlink}%
\providecommand \@sanitize@url [0]{\catcode `\\12\catcode `\$12\catcode
  `\&12\catcode `\#12\catcode `\^12\catcode `\_12\catcode `\%12\relax}%
\providecommand \@@startlink[1]{}%
\providecommand \@@endlink[0]{}%
\providecommand \url  [0]{\begingroup\@sanitize@url \@url }%
\providecommand \@url [1]{\endgroup\@href {#1}{\urlprefix }}%
\providecommand \urlprefix  [0]{URL }%
\providecommand \Eprint [0]{\href }%
\providecommand \doibase [0]{http://dx.doi.org/}%
\providecommand \selectlanguage [0]{\@gobble}%
\providecommand \bibinfo  [0]{\@secondoftwo}%
\providecommand \bibfield  [0]{\@secondoftwo}%
\providecommand \translation [1]{[#1]}%
\providecommand \BibitemOpen [0]{}%
\providecommand \bibitemStop [0]{}%
\providecommand \bibitemNoStop [0]{.\EOS\space}%
\providecommand \EOS [0]{\spacefactor3000\relax}%
\providecommand \BibitemShut  [1]{\csname bibitem#1\endcsname}%
\let\auto@bib@innerbib\@empty
\end{thebibliography}%


\begin{thebibliography}{10}
\expandafter\ifx\csname url\endcsname\relax
  \def\url#1{\texttt{#1}}\fi
\expandafter\ifx\csname urlprefix\endcsname\relax\def\urlprefix{URL }\fi
\providecommand{\bibinfo}[2]{#2}
\providecommand{\eprint}[2][]{\url{#2}}

\bibitem{Newman2010}
\bibinfo{author}{Newman, M.}
\newblock \emph{\bibinfo{title}{Networks: An Introduction}}
  (\bibinfo{publisher}{Oxford University Press, Inc.}, \bibinfo{address}{New
  York, NY, USA}, \bibinfo{year}{2010}).

\bibitem{Latora2017}
\bibinfo{author}{Latora, V.}, \bibinfo{author}{Nicosia, V.} \&
  \bibinfo{author}{Russo, G.}
\newblock \emph{\bibinfo{title}{Complex Networks: Principles, Methods and
  Applications}} (\bibinfo{publisher}{Cambridge University Press},
  \bibinfo{year}{2017}).

\bibitem{Albert2000}
\bibinfo{author}{Albert, R.}, \bibinfo{author}{Jeong, H.} \&
  \bibinfo{author}{Barab{\'a}si, A.-L.}
\newblock \bibinfo{title}{Error and attack tolerance of complex networks}.
\newblock \emph{\bibinfo{journal}{Nature}} \textbf{\bibinfo{volume}{406}},
  \bibinfo{pages}{378--382} (\bibinfo{year}{2000}).

\bibitem{USBlackout2003}
\bibinfo{author}{{New York Independent System Operator}}.
\newblock \bibinfo{title}{Interim report on the August 14, 2003, blackout}
  (\bibinfo{year}{2004}).
\newblock
  [\url{https://www.hks.harvard.edu/hepg/Papers/NYISO.blackout.report.8.Jan.04.pdf}].

\bibitem{UCTE_Report2007}
\bibinfo{author}{{Union for the Co-ordination of Transmission of Electricity
  (UCTE)}}.
\newblock \bibinfo{title}{Final report: System disturbance on 4 November 2006}
  (\bibinfo{year}{2007}).
\newblock
  [\url{https://www.entsoe.eu/fileadmin/user_upload/_library/publications/ce/otherreports/Final-Report-20070130.pdf}].

\bibitem{CERC2012}
\bibinfo{author}{{Central Electricty Regulatory Commision (CERC)}}.
\newblock \bibinfo{title}{Report on the grid disturbances on 30th July and 31st
 July 2012}.
\newblock
  [\url{http://www.cercind.gov.in/2012/orders/Final_Report_Grid_Disturbance.pdf}].

\bibitem{Bialek2007}
\bibinfo{author}{Bialek, J.~W.}
\newblock \bibinfo{title}{Why has it happened again? Comparison between the
  {UCTE} blackout in 2006 and the blackouts of 2003}.
\newblock In \emph{\bibinfo{booktitle}{Power Tech, 2007 IEEE Lausanne}},
  \bibinfo{pages}{51--56} (\bibinfo{organization}{IEEE}, \bibinfo{year}{2007}).

\bibitem{Pesc14}
\bibinfo{author}{Pesch, T.}, \bibinfo{author}{Allelein, H.-J.} \&
  \bibinfo{author}{Hake, J.-F.}
\newblock \bibinfo{title}{Impacts of the transformation of the German energy
  system on the transmission grid}.
\newblock \emph{\bibinfo{journal}{The European Physical Journal Special
  Topics}} \textbf{\bibinfo{volume}{223}}, \bibinfo{pages}{2561}
  (\bibinfo{year}{2014}).

\bibitem{Schaefer2018}
\bibinfo{author}{Sch{\"a}fer, B.}, \bibinfo{author}{Beck, C.},
  \bibinfo{author}{Aihara, K.}, \bibinfo{author}{Witthaut, D.} \&
  \bibinfo{author}{Timme, M.}
\newblock \bibinfo{title}{Non-gaussian power grid frequency fluctuations
  characterized by l{\'e}vy-stable laws and superstatistics}.
\newblock \emph{\bibinfo{journal}{Nature Energy}} \bibinfo{pages}{1}
  (\bibinfo{year}{2018}).

\bibitem{Simonsen2008}
\bibinfo{author}{Simonsen, I.}, \bibinfo{author}{Buzna, L.},
  \bibinfo{author}{Peters, K.}, \bibinfo{author}{Bornholdt, S.} \&
  \bibinfo{author}{Helbing, D.}
\newblock \bibinfo{title}{Transient dynamics increasing network vulnerability
  to cascading failures}.
\newblock \emph{\bibinfo{journal}{Physical Review Letters}}
  \textbf{\bibinfo{volume}{100}}, \bibinfo{pages}{218701}
  (\bibinfo{year}{2008}).

\bibitem{Hines2010}
\bibinfo{author}{Hines, P.}, \bibinfo{author}{Cotilla-Sanchez, E.} \&
  \bibinfo{author}{Blumsack, S.}
\newblock \bibinfo{title}{Do topological models provide good information about
  electricity infrastructure vulnerability?}
\newblock \emph{\bibinfo{journal}{Chaos: An Interdisciplinary Journal of
  Nonlinear Science}} \textbf{\bibinfo{volume}{20}}, \bibinfo{pages}{033122}
  (\bibinfo{year}{2010}).

\bibitem{Brum13}
\bibinfo{author}{Brummitt, C.~D.}, \bibinfo{author}{Hines, P. D.~H.},
  \bibinfo{author}{Dobson, I.}, \bibinfo{author}{Moore, C.} \&
  \bibinfo{author}{D'Souza, R.~M.}
\newblock \bibinfo{title}{Transdisciplinary electric power grid science}.
\newblock \emph{\bibinfo{journal}{Proceedings of the National Academy of
  Sciences}} \textbf{\bibinfo{volume}{110}}, \bibinfo{pages}{12159}
  (\bibinfo{year}{2013}).

\bibitem{Pahwa2014}
\bibinfo{author}{Pahwa, S.}, \bibinfo{author}{Scoglio, C.} \&
  \bibinfo{author}{Scala, A.}
\newblock \bibinfo{title}{Abruptness of cascade failures in power grids}.
\newblock \emph{\bibinfo{journal}{Scientific Reports}}
  \textbf{\bibinfo{volume}{4}}, \bibinfo{pages}{3694} (\bibinfo{year}{2014}).

\bibitem{Witthaut2015}
\bibinfo{author}{Witthaut, D.} \& \bibinfo{author}{Timme, M.}
\newblock \bibinfo{title}{Nonlocal effects and countermeasures in cascading
  failures}.
\newblock \emph{\bibinfo{journal}{Physical Review E}}
  \textbf{\bibinfo{volume}{92}}, \bibinfo{pages}{032809}
  (\bibinfo{year}{2015}).

\bibitem{Plietzsch2016}
\bibinfo{author}{Plietzsch, A.}, \bibinfo{author}{Schultz, P.},
  \bibinfo{author}{Heitzig, J.} \& \bibinfo{author}{Kurths, J.}
\newblock \bibinfo{title}{Local vs. global redundancy--Trade-offs between
  resilience against cascading failures and frequency stability}.
\newblock \emph{\bibinfo{journal}{The European Physical Journal Special
  Topics}} \textbf{\bibinfo{volume}{225}}, \bibinfo{pages}{551--568}
  (\bibinfo{year}{2016}).

\bibitem{Rohden2016}
\bibinfo{author}{Rohden, M.}, \bibinfo{author}{Jung, D.},
  \bibinfo{author}{Tamrakar, S.} \& \bibinfo{author}{Kettemann, S.}
\newblock \bibinfo{title}{Cascading failures in {AC} electricity grids}.
\newblock \emph{\bibinfo{journal}{Physical Review E}}
  \textbf{\bibinfo{volume}{94}}, \bibinfo{pages}{032209}
  (\bibinfo{year}{2016}).
\newblock [\url{http://link.aps.org/doi/10.1103/PhysRevE.94.032209}].

\bibitem{Manik2017}
\bibinfo{author}{Manik, D.} \emph{et~al.}
\newblock \bibinfo{title}{Network susceptibilities: Theory and applications}.
\newblock \emph{\bibinfo{journal}{Physical Review E}}
  \textbf{\bibinfo{volume}{95}}, \bibinfo{pages}{012319}
  (\bibinfo{year}{2017}).

\bibitem{Ronellenfitsch2017}
\bibinfo{author}{Ronellenfitsch, H.}, \bibinfo{author}{Manik, D.},
  \bibinfo{author}{Horsch, J.}, \bibinfo{author}{Brown, T.} \&
  \bibinfo{author}{Witthaut, D.}
\newblock \bibinfo{title}{Dual theory of transmission line outages}.
\newblock \emph{\bibinfo{journal}{IEEE Transactions on Power Systems}}
  \textbf{\bibinfo{volume}{PP}}, \bibinfo{pages}{1--1} (\bibinfo{year}{2017}).

\bibitem{Callaway2000}
\bibinfo{author}{Callaway, D.~S.}, \bibinfo{author}{Newman, M.~E.},
  \bibinfo{author}{Strogatz, S.~H.} \& \bibinfo{author}{Watts, D.~J.}
\newblock \bibinfo{title}{Network robustness and fragility: Percolation on
  random graphs}.
\newblock \emph{\bibinfo{journal}{Physical Review Letters}}
  \textbf{\bibinfo{volume}{85}}, \bibinfo{pages}{5468} (\bibinfo{year}{2000}).

\bibitem{Lozano2012}
\bibinfo{author}{Lozano, S.}, \bibinfo{author}{Buzna, L.} \&
  \bibinfo{author}{D{\'\i}az-Guilera, A.}
\newblock \bibinfo{title}{Role of network topology in the synchronization of
  power systems}.
\newblock \emph{\bibinfo{journal}{The European Physical Journal B}}
  \textbf{\bibinfo{volume}{85}}, \bibinfo{pages}{1--8} (\bibinfo{year}{2012}).

\bibitem{Albert2004}
\bibinfo{author}{Albert, R.}, \bibinfo{author}{Albert, I.} \&
  \bibinfo{author}{Nakarado, G.~L.}
\newblock \bibinfo{title}{Structural vulnerability of the North American power
  grid}.
\newblock \emph{\bibinfo{journal}{Physical Review E}}
  \textbf{\bibinfo{volume}{69}}, \bibinfo{pages}{025103}
  (\bibinfo{year}{2004}).

\bibitem{Boccaletti2006}
\bibinfo{author}{Boccaletti, S.}, \bibinfo{author}{Latora, V.},
  \bibinfo{author}{Moreno, Y.}, \bibinfo{author}{Chavez, M.} \&
  \bibinfo{author}{Hwang, D.-U.}
\newblock \bibinfo{title}{Complex networks: Structure and dynamics}.
\newblock \emph{\bibinfo{journal}{Physics Reports}}
  \textbf{\bibinfo{volume}{424}}, \bibinfo{pages}{175--308}
  (\bibinfo{year}{2006}).

\bibitem{Buldyrev2010}
\bibinfo{author}{Buldyrev, S.~V.}, \bibinfo{author}{Parshani, R.},
  \bibinfo{author}{Paul, G.}, \bibinfo{author}{Stanley, H.~E.} \&
  \bibinfo{author}{Havlin, S.}
\newblock \bibinfo{title}{Catastrophic cascade of failures in interdependent
  networks}.
\newblock \emph{\bibinfo{journal}{Nature}} \textbf{\bibinfo{volume}{464}},
  \bibinfo{pages}{1025--1028} (\bibinfo{year}{2010}).

\bibitem{Boccaletti2014}
\bibinfo{author}{Boccaletti, S.} \emph{et~al.}
\newblock \bibinfo{title}{The structure and dynamics of multilayer networks}.
\newblock \emph{\bibinfo{journal}{Physics Reports}}
  \textbf{\bibinfo{volume}{544}}, \bibinfo{pages}{1} (\bibinfo{year}{2014}).

\bibitem{Battiston2014}
\bibinfo{author}{Battiston, F.}, \bibinfo{author}{Nicosia, V.} \&
  \bibinfo{author}{Latora, V.}
\newblock \bibinfo{title}{Structural measures for multiplex networks}.
\newblock \emph{\bibinfo{journal}{Physical Review E}}
  \textbf{\bibinfo{volume}{89}}, \bibinfo{pages}{032804}
  (\bibinfo{year}{2014}).

\bibitem{Scala2016}
\bibinfo{author}{Scala, A.}, \bibinfo{author}{Lucentini, P. G. D.~S.},
  \bibinfo{author}{Caldarelli, G.} \& \bibinfo{author}{D\'Agostino, G.}
\newblock \bibinfo{title}{Cascades in interdependent flow networks}.
\newblock \emph{\bibinfo{journal}{Physica D: Nonlinear Phenomena}}
  \textbf{\bibinfo{volume}{323}}, \bibinfo{pages}{35--39}
  (\bibinfo{year}{2016}).

\bibitem{Crucitti2004}
\bibinfo{author}{Crucitti, P.}, \bibinfo{author}{Latora, V.} \&
  \bibinfo{author}{Marchiori, M.}
\newblock \bibinfo{title}{A topological analysis of the Italian electric power
  grid}.
\newblock \emph{\bibinfo{journal}{Physica A: Statistical mechanics and its
  applications}} \textbf{\bibinfo{volume}{338}}, \bibinfo{pages}{92--97}
  (\bibinfo{year}{2004}).

\bibitem{Crucitti2004a}
\bibinfo{author}{Crucitti, P.}, \bibinfo{author}{Latora, V.} \&
  \bibinfo{author}{Marchiori, M.}
\newblock \bibinfo{title}{Model for cascading failures in complex networks}.
\newblock \emph{\bibinfo{journal}{Physical Review E}}
  \textbf{\bibinfo{volume}{69}}, \bibinfo{pages}{045104}
  (\bibinfo{year}{2004}).

\bibitem{Kinney2005}
\bibinfo{author}{Kinney, R.}, \bibinfo{author}{Crucitti, P.},
  \bibinfo{author}{Albert, R.} \& \bibinfo{author}{Latora, V.}
\newblock \bibinfo{title}{Modeling cascading failures in the North American
  power grid}.
\newblock \emph{\bibinfo{journal}{The European Physical Journal B-Condensed
  Matter and Complex Systems}} \textbf{\bibinfo{volume}{46}},
  \bibinfo{pages}{101--107} (\bibinfo{year}{2005}).

\bibitem{Ji2016}
\bibinfo{author}{Ji, C.} \emph{et~al.}
\newblock \bibinfo{title}{Large-scale data analysis of power grid resilience
  across multiple US service regions}.
\newblock \emph{\bibinfo{journal}{Nature Energy}} \textbf{\bibinfo{volume}{1}},
  \bibinfo{pages}{16052} (\bibinfo{year}{2016}).

\bibitem{12powergrid}
\bibinfo{author}{Rohden, M.}, \bibinfo{author}{Sorge, A.},
  \bibinfo{author}{Timme, M.} \& \bibinfo{author}{Witthaut, D.}
\newblock \bibinfo{title}{Self-organized synchronization in decentralized power
  grids}.
\newblock \emph{\bibinfo{journal}{Physical Review Letters}}
  \textbf{\bibinfo{volume}{109}}, \bibinfo{pages}{064101}
  (\bibinfo{year}{2012}).

\bibitem{13powerlong}
\bibinfo{author}{Rohden, M.}, \bibinfo{author}{Sorge, A.},
  \bibinfo{author}{Witthaut, D.} \& \bibinfo{author}{Timme, M.}
\newblock \bibinfo{title}{Impact of network topology on synchrony of
  oscillatory power grids}.
\newblock \emph{\bibinfo{journal}{Chaos}} \textbf{\bibinfo{volume}{24}},
  \bibinfo{pages}{013123} (\bibinfo{year}{2014}).

\bibitem{14bifurcation}
\bibinfo{author}{Manik, D.} \emph{et~al.}
\newblock \bibinfo{title}{Supply networks: Instabilities without overload}.
\newblock \emph{\bibinfo{journal}{The European Physical Journal Special
  Topics}} \textbf{\bibinfo{volume}{223}}, \bibinfo{pages}{2527}
  (\bibinfo{year}{2014}).

\bibitem{Nish15}
\bibinfo{author}{Nishikawa, T.} \& \bibinfo{author}{Motter, A.~E.}
\newblock \bibinfo{title}{Comparative analysis of existing models for
  power-grid synchronization}.
\newblock \emph{\bibinfo{journal}{New Journal of Physics}}
  \textbf{\bibinfo{volume}{17}}, \bibinfo{pages}{015012}
  (\bibinfo{year}{2015}).

\bibitem{16critical}
\bibinfo{author}{Witthaut, D.}, \bibinfo{author}{Rohden, M.},
  \bibinfo{author}{Zhang, X.}, \bibinfo{author}{Hallerberg, S.} \&
  \bibinfo{author}{Timme, M.}
\newblock \bibinfo{title}{Critical links and nonlocal rerouting in complex
  supply networks}.
\newblock \emph{\bibinfo{journal}{Physical Review Letters}}
  \textbf{\bibinfo{volume}{116}}, \bibinfo{pages}{138701}
  (\bibinfo{year}{2016}).

\bibitem{Kund94}
\bibinfo{author}{Kundur, P.}, \bibinfo{author}{Balu, N.~J.} \&
  \bibinfo{author}{Lauby, M.~G.}
\newblock \emph{\bibinfo{title}{Power system stability and control}},
  vol.~\bibinfo{volume}{7} (\bibinfo{publisher}{McGraw-hill New York},
  \bibinfo{year}{1994}).

\bibitem{Pourbeik2006}
\bibinfo{author}{Pourbeik, P.}, \bibinfo{author}{Kundur, P.~S.} \&
  \bibinfo{author}{Taylor, C.~W.}
\newblock \bibinfo{title}{The anatomy of a power grid blackout-root causes and
  dynamics of recent major blackouts}.
\newblock \emph{\bibinfo{journal}{IEEE Power and Energy Magazine}}
  \textbf{\bibinfo{volume}{4}}, \bibinfo{pages}{22--29} (\bibinfo{year}{2006}).

\bibitem{Yang2017b}
\bibinfo{author}{Yang, Y.} \& \bibinfo{author}{Motter, A.~E.}
\newblock \bibinfo{title}{Cascading failures as continuous phase-space
  transitions}.
\newblock \emph{\bibinfo{journal}{Physical Review Letters}}
  \textbf{\bibinfo{volume}{119}}, \bibinfo{pages}{248302}
  (\bibinfo{year}{2017}).

\bibitem{Bienstock2011}
\bibinfo{author}{Bienstock, D.}
\newblock \bibinfo{title}{Optimal control of cascading power grid failures}.
\newblock In \emph{\bibinfo{booktitle}{2011 50th IEEE conference on Decision
  and control and European control conference (CDC-ECC)}},
  \bibinfo{pages}{2166--2173} (\bibinfo{organization}{IEEE},
  \bibinfo{year}{2011}).

\bibitem{Brockmann2013}
\bibinfo{author}{Brockmann, D.} \& \bibinfo{author}{Helbing, D.}
\newblock \bibinfo{title}{The hidden geometry of complex, network-driven
  contagion phenomena}.
\newblock \emph{\bibinfo{journal}{Science}} \textbf{\bibinfo{volume}{342}},
  \bibinfo{pages}{1337--1342} (\bibinfo{year}{2013}).

\bibitem{Yang2017a}
\bibinfo{author}{Yang, Y.}, \bibinfo{author}{Nishikawa, T.} \&
  \bibinfo{author}{Motter, A.~E.}
\newblock \bibinfo{title}{Small vulnerable sets determine large network
  cascades in power grids}.
\newblock \emph{\bibinfo{journal}{Science}} \textbf{\bibinfo{volume}{358}},
  \bibinfo{pages}{eaan3184} (\bibinfo{year}{2017}).

\bibitem{05vuln}
\bibinfo{author}{Latora, V.} \& \bibinfo{author}{Marchiori, M.}
\newblock \bibinfo{title}{Vulnerability and protection of infrastructure
  networks}.
\newblock \emph{\bibinfo{journal}{Physical Review E}}
  \textbf{\bibinfo{volume}{71}}, \bibinfo{pages}{015103}
  (\bibinfo{year}{2005}).

\bibitem{PhysRevLett.93.068701}
\bibinfo{author}{Argollo~de Menezes, M.} \& \bibinfo{author}{Barab\'asi, A.-L.}
\newblock \bibinfo{title}{Separating internal and external dynamics of complex
  systems}.
\newblock \emph{\bibinfo{journal}{Physical Review Letters}}
  \textbf{\bibinfo{volume}{93}}, \bibinfo{pages}{068701}
  (\bibinfo{year}{2004}).

\bibitem{08Meloni}
\bibinfo{author}{Meloni, S.}, \bibinfo{author}{G\'omez-Garde\~nes, J.},
  \bibinfo{author}{Latora, V.} \& \bibinfo{author}{Moreno, Y.}
\newblock \bibinfo{title}{Scaling breakdown in flow fluctuations on complex
  networks}.
\newblock \emph{\bibinfo{journal}{Physical Review Letters}}
  \textbf{\bibinfo{volume}{100}}, \bibinfo{pages}{208701}
  (\bibinfo{year}{2008}).

\bibitem{ccolak2016understanding}
\bibinfo{author}{{\c{C}}olak, S.}, \bibinfo{author}{Lima, A.} \&
  \bibinfo{author}{Gonz{\'a}lez, M.~C.}
\newblock \bibinfo{title}{Understanding congested travel in urban areas}.
\newblock \emph{\bibinfo{journal}{Nature Communications}}
  \textbf{\bibinfo{volume}{7}}, \bibinfo{pages}{10793} (\bibinfo{year}{2016}).

\bibitem{lima2016understanding}
\bibinfo{author}{Lima, A.}, \bibinfo{author}{Stanojevic, R.},
  \bibinfo{author}{Papagiannaki, D.}, \bibinfo{author}{Rodriguez, P.} \&
  \bibinfo{author}{Gonz{\'a}lez, M.~C.}
\newblock \bibinfo{title}{Understanding individual routing behaviour}.
\newblock \emph{\bibinfo{journal}{Journal of The Royal Society Interface}}
  \textbf{\bibinfo{volume}{13}}, \bibinfo{pages}{20160021}
  (\bibinfo{year}{2016}).

\bibitem{Echenique2005}
\bibinfo{author}{Echenique, P.}, \bibinfo{author}{G{\'o}mez-Garde{\~{n}}es, J.}
  \& \bibinfo{author}{Moreno, Y.}
\newblock \bibinfo{title}{Dynamics of jamming transitions in complex networks}.
\newblock \emph{\bibinfo{journal}{Europhysics Letters}}
  \textbf{\bibinfo{volume}{71}}, \bibinfo{pages}{325} (\bibinfo{year}{2005}).

\bibitem{Petrone2016}
\bibinfo{author}{Petrone, D.} \& \bibinfo{author}{Latora, V.}
\newblock \bibinfo{title}{A hybrid approach to assess systemic risk in
  financial networks}.
\newblock \emph{\bibinfo{journal}{arXiv preprint arXiv:1610.00795}}
  (\bibinfo{year}{2016}).

\bibitem{Wood13}
\bibinfo{author}{Wood, A.~J.}, \bibinfo{author}{Wollenberg, B.~F.} \&
  \bibinfo{author}{Shebl\'e, G.~B.}
\newblock \emph{\bibinfo{title}{Power Generation, Operation and Control}}
  (\bibinfo{publisher}{John Wiley \& Sons}, \bibinfo{address}{New York},
  \bibinfo{year}{2013}).

\bibitem{Mach08}
\bibinfo{author}{Machowski, J.}, \bibinfo{author}{Bialek, J.} \&
  \bibinfo{author}{Bumby, J.}
\newblock \emph{\bibinfo{title}{Power system dynamics, stability and control}}
  (\bibinfo{publisher}{John Wiley \& Sons}, \bibinfo{address}{New York},
  \bibinfo{year}{2008}).

\bibitem{Motter2002}
\bibinfo{author}{Motter, A.~E.} \& \bibinfo{author}{Lai, Y.-C.}
\newblock \bibinfo{title}{Cascade-based attacks on complex networks}.
\newblock \emph{\bibinfo{journal}{Physical Review E}}
  \textbf{\bibinfo{volume}{66}}, \bibinfo{pages}{065102}
  (\bibinfo{year}{2002}).

\bibitem{Koc2014}
\bibinfo{author}{Ko{\c{c}}, Y.}, \bibinfo{author}{Warnier, M.},
  \bibinfo{author}{Van~Mieghem, P.}, \bibinfo{author}{Kooij, R.~E.} \&
  \bibinfo{author}{Brazier, F.~M.}
\newblock \bibinfo{title}{A topological investigation of phase transitions of
  cascading failures in power grids}.
\newblock \emph{\bibinfo{journal}{Physica A: Statistical Mechanics and its
  Applications}} \textbf{\bibinfo{volume}{415}}, \bibinfo{pages}{273--284}
  (\bibinfo{year}{2014}).

\bibitem{Rosato2007}
\bibinfo{author}{Rosato, V.}, \bibinfo{author}{Bologna, S.} \&
  \bibinfo{author}{Tiriticco, F.}
\newblock \bibinfo{title}{Topological properties of high-voltage electrical
  transmission networks}.
\newblock \emph{\bibinfo{journal}{Electric Power Systems Research}}
  \textbf{\bibinfo{volume}{77}}, \bibinfo{pages}{99--105}
  (\bibinfo{year}{2007}).

\bibitem{Manik2016a}
\bibinfo{author}{Manik, D.}, \bibinfo{author}{Timme, M.} \&
  \bibinfo{author}{Witthaut, D.}
\newblock \bibinfo{title}{Cycle flows and multistability in oscillatory
  networks}.
\newblock \emph{\bibinfo{journal}{Chaos: An Interdisciplinary Journal of
  Nonlinear Science}} \textbf{\bibinfo{volume}{27}}, \bibinfo{pages}{083123}
  (\bibinfo{year}{2017}).

\bibitem{12klemm}
\bibinfo{author}{K.~Klemm, V. M.~E., M. \'A.~Serrano} \&
  \bibinfo{author}{Miguel, M.~S.}
\newblock \bibinfo{title}{A measure of individual role in collective dynamics}.
\newblock \emph{\bibinfo{journal}{Scientific Reports}}
  \textbf{\bibinfo{volume}{2}}, \bibinfo{pages}{292} (\bibinfo{year}{2012}).

\bibitem{Hines2008}
\bibinfo{author}{Hines, P.} \& \bibinfo{author}{Blumsack, S.}
\newblock \bibinfo{title}{A centrality measure for electrical networks}.
\newblock In \emph{\bibinfo{booktitle}{Hawaii International Conference on
  System Sciences, Proceedings of the 41st Annual}}, \bibinfo{pages}{185--185}
  (\bibinfo{organization}{IEEE}, \bibinfo{year}{2008}).

\bibitem{Hines2017}
\bibinfo{author}{Hines, P.~D.}, \bibinfo{author}{Dobson, I.} \&
  \bibinfo{author}{Rezaei, P.}
\newblock \bibinfo{title}{Cascading power outages propagate locally in an
  influence graph that is not the actual grid topology}.
\newblock \emph{\bibinfo{journal}{IEEE Transactions on Power Systems}}
  \textbf{\bibinfo{volume}{32}}, \bibinfo{pages}{958--967}
  (\bibinfo{year}{2017}).

\bibitem{Simpson-Porco2016}
\bibinfo{author}{Simpson-Porco, J.~W.}, \bibinfo{author}{D{\"o}rfler, F.} \&
  \bibinfo{author}{Bullo, F.}
\newblock \bibinfo{title}{Voltage collapse in complex power grids}.
\newblock \emph{\bibinfo{journal}{Nature communications}}
  \textbf{\bibinfo{volume}{7}} (\bibinfo{year}{2016}).

\bibitem{Auer2016}
\bibinfo{author}{Auer, S.}, \bibinfo{author}{Kleis, K.},
  \bibinfo{author}{Schultz, P.}, \bibinfo{author}{Kurths, J.} \&
  \bibinfo{author}{Hellmann, F.}
\newblock \bibinfo{title}{The impact of model detail on power grid resilience
  measures}.
\newblock \emph{\bibinfo{journal}{The European Physical Journal Special
  Topics}} \textbf{\bibinfo{volume}{225}}, \bibinfo{pages}{609--625}
  (\bibinfo{year}{2016}).

\bibitem{Schmietendorf2014}
\bibinfo{author}{Schmietendorf, K.}, \bibinfo{author}{Peinke, J.},
  \bibinfo{author}{Friedrich, R.} \& \bibinfo{author}{Kamps, O.}
\newblock \bibinfo{title}{Self-organized synchronization and voltage stability
  in networks of synchronous machines}.
\newblock \emph{\bibinfo{journal}{The European Physical Journal Special
  Topics}} \textbf{\bibinfo{volume}{223}}, \bibinfo{pages}{2577--2592}
  (\bibinfo{year}{2014}).

\bibitem{Sharafutdinov2017}
\bibinfo{author}{Sharafutdinov, K.}, \bibinfo{author}{Matthiae, M.},
  \bibinfo{author}{Faulwasser, T.} \& \bibinfo{author}{Witthaut, D.}
\newblock \bibinfo{title}{Rotor-angle versus voltage instability in the
  third-order model}.
\newblock \emph{\bibinfo{journal}{arXiv preprint arXiv:1706.06396}}
  (\bibinfo{year}{2017}).

\bibitem{Ma2016}
\bibinfo{author}{Ma, J.}, \bibinfo{author}{Sun, Y.}, \bibinfo{author}{Yuan,
  X.}, \bibinfo{author}{Kurths, J.} \& \bibinfo{author}{Zhan, M.}
\newblock \bibinfo{title}{Dynamics and collapse in a power system model with
  voltage variation: The damping effect}.
\newblock \emph{\bibinfo{journal}{PloS one}} \textbf{\bibinfo{volume}{11}},
  \bibinfo{pages}{e0165943} (\bibinfo{year}{2016}).

\bibitem{Cetinay2017}
\bibinfo{author}{Cetinay, H.}, \bibinfo{author}{Soltan, S.},
  \bibinfo{author}{Kuipers, F.~A.}, \bibinfo{author}{Zussman, G.} \&
  \bibinfo{author}{Van~Mieghem, P.}
\newblock \bibinfo{title}{Comparing the effects of failures in power grids
  under the AC and DC power flow models}.
\newblock \emph{\bibinfo{journal}{IEEE Transactions on Network Science and
  Engineering}}  (\bibinfo{year}{2017}).

\bibitem{Salmeron2004}
\bibinfo{author}{Salmeron, J.}, \bibinfo{author}{Wood, K.} \&
  \bibinfo{author}{Baldick, R.}
\newblock \bibinfo{title}{Analysis of electric grid security under terrorist
  threat}.
\newblock \emph{\bibinfo{journal}{IEEE Transactions on Power Systems}}
  \textbf{\bibinfo{volume}{19}}, \bibinfo{pages}{905--912}
  (\bibinfo{year}{2004}).

\bibitem{Fila08}
\bibinfo{author}{Filatrella, G.}, \bibinfo{author}{Nielsen, A.~H.} \&
  \bibinfo{author}{Pedersen, N.~F.}
\newblock \bibinfo{title}{Analysis of a power grid using a Kuramoto-like
  model}.
\newblock \emph{\bibinfo{journal}{The European Physical Journal B}}
  \textbf{\bibinfo{volume}{61}}, \bibinfo{pages}{485} (\bibinfo{year}{2008}).

\bibitem{entsoe14}
\bibinfo{author}{{European Network of Transmission System Operators for
  Electricity (ENTSO-E)}}.
\newblock \bibinfo{title}{Statistical factsheet 2014}.
\newblock
  \bibinfo{howpublished}{https://www.entsoe.eu/publications/major-publications/Pages/default.aspx}.
\newblock \bibinfo{note}{Accessed: 2015-09-01}.

\bibitem{Pikovsky2003}
\bibinfo{author}{Pikovsky, A.}, \bibinfo{author}{Rosenblum, M.} \&
  \bibinfo{author}{Kurths, J.}
\newblock \emph{\bibinfo{title}{Synchronization: a universal concept in
  nonlinear sciences}} (\bibinfo{publisher}{Cambridge University Press},
  \bibinfo{year}{2003}).

\bibitem{Kura75}
\bibinfo{author}{Kuramoto, Y.}
\newblock \bibinfo{title}{Self-entrainment of a population of coupled
  non-linear oscillators}.
\newblock In \bibinfo{editor}{Araki, H.} (ed.)
  \emph{\bibinfo{booktitle}{International Symposium on on Mathematical Problems
  in Theoretical Physics}}, Lecture Notes in Physics Vol. 39,
  \bibinfo{pages}{420} (\bibinfo{publisher}{Springer}, \bibinfo{address}{New
  York}, \bibinfo{year}{1975}).

\bibitem{Stro00}
\bibinfo{author}{Strogatz, S.~H.}
\newblock \bibinfo{title}{From {Kuramoto} to {Crawford}: Exploring the onset of
  synchronization in populations of coupled oscillators}.
\newblock \emph{\bibinfo{journal}{Physica D: Nonlinear Phenomena}}
  \textbf{\bibinfo{volume}{143}}, \bibinfo{pages}{1} (\bibinfo{year}{2000}).

\bibitem{Witthaut2017}
\bibinfo{author}{Witthaut, D.}, \bibinfo{author}{Wimberger, S.},
  \bibinfo{author}{Burioni, R.} \& \bibinfo{author}{Timme, M.}
\newblock \bibinfo{title}{Classical synchronization indicates persistent
  entanglement in isolated quantum systems}.
\newblock \emph{\bibinfo{journal}{Nature Communications}}
  \textbf{\bibinfo{volume}{8}} (\bibinfo{year}{2017}).

\end{thebibliography}
\end{document}